\documentclass[10pt]{iopart}

\usepackage{iopams}
\usepackage{amssymb}
\usepackage{amsfonts}
\usepackage{amstext}
\usepackage{graphicx}
\usepackage{setstack}
\usepackage{amscd}
\usepackage{mathrsfs}
\usepackage{stmaryrd}
\usepackage{slashed}
\usepackage{color}

\newcommand{\Orb}{\mathrm{Orb}}
\newcommand{\ad}{\mathrm{ad}}
\newcommand{\Ad}{\mathrm{Ad}}
\newcommand{\llangle}{\langle \hspace{-0.2em} \langle}
\newcommand{\rrangle}{\rangle \hspace{-0.2em} \rangle}
\newcommand{\Sp}{\mathrm{Sp}}
\newcommand{\mod}{\, \mathrm {mod}\,}

\begin{document}

\title{Effective Hamiltonians for almost-periodically driven quantum systems}

\author{David Viennot}
\address{Institut UTINAM (CNRS UMR 6213, Universit\'e de Bourgogne-Franche-Comt\'e, Observatoire de Besan\c con), 41bis Avenue de l'Observatoire, BP1615, 25010 Besan\c con cedex, France.}

\begin{abstract}
We present an effective Hamiltonian theory available for some quasi-periodically driven quantum systems which does not need the knowledge of the Fourier frequencies of the control signal. It could also be available for some chaotically driven quantum systems. It is based on the Koopman approach which generalizes the Floquet approach used with periodically driven systems. We show the properties of the quasi-energy states (eigenvectors of the effective Hamiltonian) as quasi-recurrent states of the quantum system.
\end{abstract}



\section{Introduction}
Periodically driven quantum systems is a subject of great interest in quantum physics. It is well known that the consistent framework to treat this subject is the Floquet theorem \cite{Floquet} which has been firstly considered in quantum dynamics in \cite{Shirley}. Since this pionner work, the subject has been extensively studied \cite{Sambe,Barone,Moore1,Haake,Moore2,Guerin1,Guerin2,Drese,Guerin3,Miyamoto,Viennot1,Else}. A usefull method to study periodically driven quantum systems, which is directly induced by the Floquet theory, consists to use a time-independent effectif Hamiltonian governing the stroboscopic dynamics (evolution on the whole period) \cite{Goldman}.\\
Some attempts to generalize this approach has been proposed for quasi-periodic driven systems (time-dependent systems characterized by several irrationally related frequencies) \cite{Guerin3,Neumann,Verdeny1,Verdeny2}. These studies has been confronted to the fact the Floquet theory cannot be applied for non-periodic systems. In this paper, we want generalize the effective Hamiltonian approach to almost-periodically driven systems, i.e. systems such that $\forall t, \forall \epsilon>0$, $\exists T_{\epsilon,t}>\eta_{\epsilon,t}$ such that $\|H(t+T_{\epsilon,t}) - H(t)\| < \epsilon$, with $\eta_{\epsilon,t}$ such that $\|H(t+\eta_{\epsilon,t})-H(t)\|>\epsilon$; $H(t)$ being the time-dependent Hamiltonian of the driven system. This situation includes periodically and quasi-periodically driven systems (but in contrast with the previous works, we do not need the decomposition of time-dependent Hamiltonian into Fourier modes associated with each frequencies), but also systems driven by classical flows with Poincar\'e recurrence \cite{Lasota} (including chaotic Hamiltonian flows) and systems driven by some stochastic flows as for example Brownian motions onto a compact manifold without boundary. These two situations can modelize a quantum system driven by a periodic control but affected by (chaotic or stochastic) noises \cite{Viennot2}. Our approach is based on the Koopman approach of the dynamical systems \cite{Lasota,Koopman1,Koopman2} and follows our previous work concerning the mathematical properties of the Schr\"odinger-Koopman quasienergy states \cite{Viennot3}. In some sense, our appoach generalizes to time-dependent systems the phenomenon of quantum recurrence/revival \cite{Bocchieri,Eberly} found for time-independent Hamiltonians.\\
This paper is organized as follows. Section 2 introduces the effective Hamiltonians governing almost-periodically driven quantum systems. Concrete formulae for these effective Hamiltonians are computed in section 3 and we present the expected dynamical behaviours induced by the almost-periodicity. Finally, some illustrations are presented section 4.

\section{SK and first recurrence effective Hamiltonians}
\subsection{The generic model}
We consider a quantum system described by a Hilbert space $\mathcal H$ and governed for its free evolution by a Hamiltonian $\hbar \omega_1 \hat H$ where $\hbar \omega_1$ is the characteristic transition energy of the system ($\hat H$ is the reduced free Hamiltonian). The quantum system is driven by a classical discrete flow $\varphi: \Gamma \to \Gamma$ onto a phase space $\Gamma$ supposed to be a compact manifold without boundary (in general $\Gamma$ is a $N$-torus). Let $\mu: \mathscr T \to \mathbb R^+$ be an invariant measure onto $\Gamma$ ($\mathscr T$ is a $\sigma$-algebra of $\Gamma$ (generally the Borelian $\sigma$-algebra), and $\forall O \in \mathscr T$ (open set of $\Gamma$), $\mu(\varphi(O)) = \mu(O)$), and such that $\mu(\Gamma)< \infty$. Let $T_0$ be the sampling period on which the dynamics of the quantum system is discretized. We suppose that the evolution operator of the driven quantum system can be written in the following form $\forall n \in \mathbb N$:
\begin{equation}
  U_n \equiv U((n+1)T_0,nT_0)=e^{-\imath \frac{\omega_1}{\omega_0} \hat H} e^{-\imath V(\varphi^n(\theta_0))}
\end{equation}
where $\omega_0=\frac{2\pi}{T_0}$ is the sampling frequency and $V(\theta)$ is the interaction operator for the value $\theta \in \Gamma$ of the control parameters, $\theta_0 \in \Gamma$ are the initial values of the control parameters. This form is very general. It can correspond to a time-dependent Hamiltonian $H(t) = H_0+V(\varphi^{t}(\theta_0))$ where $\varphi^{t}(\theta_0) = \theta(t)$ are continuous time-dependent parameters. With $T_0 \ll \frac{2\pi}{\hbar \omega_1}$ we have $U_n = e^{-\imath \frac{\omega_1}{\omega_0} \hat H} e^{-\imath V(\varphi^{nT_0}(\theta_0))} + \mathcal O\left(\frac{\omega_1}{\omega_0}\right)$ (with $\hat H = \frac{H_0}{\hbar \omega_1}$). It can also correspond to the time-dependent Hamiltonian $H(t)=H_0+\sum_{n \in \mathbb N} W(\varphi^n(\theta_0)) \delta\left(t-nT_0+\frac{\Delta(\varphi^n(\theta_0))}{\omega_0}\right)$ of a kicked quantum system, where $W(\theta)$ is the kick operator for the values $\theta \in \Gamma$ of the control parameters and $0 \leq \Delta(\theta)<2\pi$ is the ``angular'' delay of the kick for the value $\theta$ (the quantum system is kicked once during a period $T_0$ but the kick can be delayed). In that case, $U_n = e^{-\imath \frac{H_0}{\hbar \omega_0} (2\pi - \Delta(\varphi^n(\theta_0)))} e^{-\imath W(\varphi^n(\theta_0))} e^{-\imath \frac{H_0}{\hbar \omega_0} \Delta(\varphi^n(\theta_0))}$ (see for example \cite{Viennot1}), which can be rewritten as $U_n=e^{-\imath \frac{\omega_1}{\omega_0} \hat H} e^{-\imath V(\varphi^n(\theta_0))}$ with $\hat H = \frac{2\pi}{\hbar \omega_1} H_0$ and $V(\varphi^n(\theta_0)) = e^{\imath \frac{H_0}{\hbar \omega_0} \Delta(\varphi^n(\theta_0))} W(\varphi^n(\theta_0)) e^{-\imath \frac{H_0}{\hbar \omega_0} \Delta(\varphi^n(\theta_0))}$.\\

By the Poincar\'e recurrence theorem \cite{Lasota}, we have for $\mu$-almost all $\theta \in \Gamma$
\begin{equation} \label{poincare}
  \forall \epsilon>0, \exists p_{\epsilon,\theta}>0, \|\varphi^{p_{\epsilon,\theta}}(\theta)-\theta\|< \epsilon
\end{equation}
whereas $\exists n < p_{\epsilon,\theta}$ for which $\|\varphi^n(\theta)-\theta\|>\epsilon$ (the norm in $\Gamma$ is the Euclidean norm of the control parameters $\|\theta\|^2 = \sum_i (\theta^i)^2$). $p_{\epsilon,\theta}$ being not unique, we set $p_{\epsilon,\theta}$ as being the smallest value satisfying the relation (\ref{poincare}). If $\theta_0$ is $p$-cyclic ($\varphi^p(\theta_0)=\theta_0$), then $p_{\epsilon,\theta_0} = p_{\epsilon,\varphi^n(\theta_0)} = p$ ($p_{\epsilon,\theta}$ is independent of $\epsilon$ and is the same for all point of the orbit of $\theta_0$ : $\Orb(\theta_0) = \{\varphi^n(\theta_0)\}_{n\in \mathbb N}$). This case corresponds to a periodically driven quantum system. We recover the quasi-periodic case if $\overline{\Orb(\theta_0)}$ is a torus (the overline denotes the topological closure) and if $p_{\epsilon,\varphi^n(\theta_0)} = p_{\epsilon,\theta_0}$ (the almost-period is the same on the whole of $\Orb(\theta_0)$). For the case of a chaotic flow $\varphi$, $p_{\epsilon,\theta}$ is ``erratically'' dependent on $\epsilon$ and $\theta$. And finally, if $\varphi$ is a stochastic flow, $p_{\epsilon,\theta}$ is a random variable. Since a flow can have several behaviors, it can be interesting to decompose the phase space into ergodic components: $\Gamma = \bigcup_e \Gamma_e$, with $\mu(\Gamma_e \cap \Gamma_{e'}) = 0$ (for $e'\not=e$) and with $\Gamma_e = \overline{\Orb(\theta)}$ for $\mu$-almost all $\theta \in \Gamma_e$.\\
$\varphi$ modelizes a control applied on the quantum system (by electromagnetic fields, STM, ultra-fast kicks,...), or a classical noise affecting the quantum system (when $\varphi$ is chaotic or stochastic); or the both ones.

\subsection{Definition of the effective Hamiltonians}
Let $U(\theta) = e^{-\imath \frac{\omega_1}{\omega_0} \hat H} e^{- \imath V(\theta)}$. The Schr\"odinger-Koopman (SK) quasienergy states are defined as solutions of the equation (see \cite{Viennot3}):
\begin{equation} \label{SKquasistate}
  U(\theta) |Z\mu_{ie},\theta \rangle = e^{-\imath \chi_{ie}} |Z\mu_{ie},\varphi(\theta) \rangle
\end{equation}
for $\theta \in \Gamma_e$ and where $\chi_{ie}$ is called quasienergy ($\chi_{ie}$ depends only on the ergodic component $\Gamma_e$ on which $\theta$ belongs). A quasienergy becomes another quasienergy under the gauge change: $\chi_{ie} \to \chi_{ie} - \imath \lambda$ and $|Z\mu_{ie},\theta \rangle \to f_\lambda(\theta) |Z\mu_{ie},\theta \rangle$, where $\lambda$ and $f_\lambda$ are a Koopman value and the associated Koopman mode (i.e. $f_\lambda(\varphi(\theta)) = e^\lambda f_\lambda(\theta)$, $\lambda \in \imath \mathbb R$). $\chi_{ie}$ does not depend on $\theta \in \Gamma_e$ (due to the quasienergy orbital stability theorem \cite{Viennot3}) and in general $|Z\mu_{ie},\theta \rangle=0$ for $\theta \not\in \Gamma_e$. We choose $\{\chi_{ie}\}_{i=1,...,\dim \mathcal H; e}$ such that $(|Z\mu_{ie},\theta \rangle)_{i=1,...,\dim \mathcal H}$ be a basis of $\mathcal H$ for all $\theta \in \Gamma_e$. $\{\chi_{ie}\}_{i=1,...,\dim \mathcal H; e}$ is called fundamental quasienergy spectrum of the driven quantum system. Moreover for $\mu$-almost all $\theta \in \Gamma_e$, $\langle Z\mu_{ie},\theta|Z\mu_{je},\theta \rangle = \delta_{ij}$.\\
From equation (\ref{SKquasistate}) we define the SK effective Hamiltonian as being:
\begin{equation} \label{SKHeff}
  H^{eff}(\theta) = \sum_e \sum_i \chi_{ie} |Z\mu_{ie},\theta \rangle \langle Z\mu_{ie},\theta |
\end{equation}
(by assuming that $|Z\mu_{ie},\theta\rangle = 0$ for $\theta \not\in \Gamma_e$). To understand the role of $H^{eff}(\theta)$ it needs to consider its relation with the first recurrence Hamiltonian $H^{eff}_\epsilon(\theta)$ defined by
\begin{equation} \label{FRHeff}
  e^{-\imath p_{\epsilon,\theta} H^{eff}_\epsilon(\theta)} = U(\varphi^{p_{\epsilon,\theta}-1}(\theta))...U(\varphi(\theta))U(\theta)
\end{equation}
Firstly, if $\theta$ is $p$-cyclic, then $H^{eff}_\epsilon(\theta)$ is independent of $\epsilon$, $e^{-\imath p H^{eff}_\epsilon(\theta)} = U(\varphi^{p-1}(\theta))...U(\theta)$ and is equal to the SK effective Hamiltonian $H^{eff}_\epsilon(\theta)=H^{eff}(\theta)$ which is here the usual Floquet effective Hamiltonian of the periodic driven quantum system ($|Z\mu_{ie},\theta\rangle$ is the Floquet quasienergy state defined by $U(\varphi^{p-1}(\theta))...U(\theta)|Z\mu_{ie},\theta \rangle = e^{-\imath p \chi_{ie}} |Z\mu_i,\theta \rangle$, see \cite{Viennot1}).\\
If $\theta$ is not $p$-cyclic, we have only $U(\varphi^{p_{\epsilon,\theta}-1}(\theta))...U(\theta)|Z\mu_{ie},\theta \rangle = e^{-\imath p_{\epsilon,\theta} \chi_{ie}} |Z\mu_{ie},\varphi^{p_{\epsilon,\theta}}(\theta) \rangle$, but because of eq. (\ref{poincare}) we have
\begin{equation}
  |Z\mu_{ie},\varphi^{p_{\epsilon,\theta}}(\theta) \rangle = |Z\mu_{ie},\theta \rangle + \partial_\nu |Z\mu_{ie},\theta \rangle \tilde \epsilon^\nu(\theta) + \mathcal O(\epsilon^2)
\end{equation}
where $\tilde \epsilon(\theta) \equiv \varphi^{p_{\epsilon,\theta}}(\theta)-\theta$ ($\|\tilde \epsilon(\theta)\|=\mathcal O(\epsilon)$). Finally, we have
$e^{-\imath p_{\epsilon,\theta} H^{eff}_\epsilon(\theta)}|Z\mu_{ie},\theta \rangle = e^{-\imath p_{\epsilon,\theta} \chi_{ie}} |Z\mu_{ie},\theta\rangle + e^{-\imath p_{\epsilon,\theta} \chi_{ie}} \partial_\nu |Z\mu_{ie},\theta\rangle \tilde \epsilon^\nu(\theta) + \mathcal O(\epsilon^2)$ and then
\begin{equation}
  e^{-\imath p_{\epsilon,\theta} H^{eff}_\epsilon(\theta)} = \left(1+A_{\nu}(\theta)\tilde \epsilon^\nu(\theta) + \mathcal O(\epsilon^2) \right)e^{-\imath p_{\epsilon,\theta}H^{eff}(\theta)}
\end{equation}
with
\begin{equation}
  A_{\nu}(\theta) = \sum_e \sum_{ij} \langle Z\mu_{je},\theta|\partial_\nu|Z\mu_{ie},\theta\rangle |Z\mu_{je},\theta\rangle \langle Z\mu_{ie},\theta|
\end{equation}
More precisely, consider a sequence $(\epsilon_n)_{n\in \mathbb N}$ such that $\epsilon_{n+1}<\epsilon_n$, $\lim_{n \to +\infty} \epsilon_n=0$ and such that $\forall \epsilon \in ]\epsilon_{n+1},\epsilon_n]$, $p_{\epsilon,\theta}=p_{\epsilon_n,\theta}$. Since $\theta$ is not cyclic, $\lim_{n\to +\infty} p_{\epsilon_n,\theta} = +\infty$. We have clearly,
\begin{equation} \label{limit}
  \lim_{n \to +\infty} \|e^{\imath p_{\epsilon_n,\theta} H^{eff}(\theta)} e^{-\imath p_{\epsilon_n,\theta} H^{eff}_{\epsilon_n}(\theta)} -1 \|=0
\end{equation}
The SK effective Hamiltonian $H^{eff}$ is the limit in sense of eq. (\ref{limit}) of the first recurrence Hamiltonian when the recurrence accuracy tends to zero. Moreover, since $e^{-\imath p_{\epsilon,\theta} H^{eff}_\epsilon(\theta)} = e^{A_{\nu}(\theta)\tilde \epsilon^\nu(\theta) + \mathcal O(\epsilon^2) } e^{-\imath p_{\epsilon,\theta}H^{eff}(\theta)}$, we have (see \ref{BCH}):
\begin{equation}\label{relatHeff}
  H^{eff}_\epsilon(\theta) = H^{eff}(\theta)+\imath \mathscr A_{\nu}(\theta) \frac{\tilde \epsilon^\nu(\theta)}{p_{\epsilon,\theta}} + \mathcal O\left(\frac{\epsilon^2}{p_{\epsilon,\theta}} \right)
\end{equation}
where
\begin{equation}
  \mathscr A_{\nu}(\theta)_{eji} = \left\{\begin{array}{ll} A_{\nu}(\theta)_{eii} & \text{if $i=j$} \\ \frac{\imath p_{\epsilon,\theta}(\chi_{ie}-\chi_{je})}{1-e^{-\imath p_{\epsilon,\theta}(\chi_{ie}-\chi_{je})}} A_{\nu}(\theta)_{eji} & \text{if $i\not=j$} \end{array} \right.
  \end{equation}
(with $A_{eji} \equiv \langle Z\mu_{je},\theta|A|Z\mu_{ie},\theta\rangle$). Note that $H^{eff}$ is the limit of $H^{eff}_\epsilon$ in the sense of eq. (\ref{limit}), but $\lim_{n \to +\infty}\|H^{eff}_{\epsilon_n}(\theta)-H^{eff}(\theta)\|\not= 0$ since $(1-e^{-\imath p(\chi_{ie}-\chi_{je})})^{-1}$ has no limit at $p \to +\infty$. Note that $p_{\epsilon,\theta}$ can be very large, the mean Poincar\'e recurrence time being $\langle p_{\epsilon,\theta} \rangle \sim \frac{\mu(\Gamma_e)}{\epsilon^{\dim \Gamma}}$ (by supposing that $\mu(\mathcal B_{\epsilon}(\theta)) \propto \epsilon^{\dim \Gamma}$ where $\mathcal B_{\epsilon}(\theta)$ is the ball of radius $\epsilon$ and centered on $\theta$ in $\Gamma$).

\subsection{Perturbation of quasienergy states}
It could be interesting to relate the eigensystems of $H^{eff}$ and $H^{eff}_\epsilon$. We can compute the first recurrence eigenstates, $H^{eff}_\epsilon(\theta)|Z\mu_i,\theta,\epsilon \rangle = \chi_{ie,\epsilon}(\theta) |Z\mu_i,\theta,\epsilon \rangle$, by using a perturbative expansion from eq. (\ref{relatHeff}):
\begin{equation}
  \chi_{ie,\epsilon}(\theta) = \chi_{ie} + \imath \langle Z\mu_i,\theta|\partial_\nu|Z\mu_i,\theta \rangle \frac{\tilde \epsilon^\nu(\theta)}{p_{\epsilon,\theta}} + \mathcal O\left(\frac{\epsilon^2}{p_{\epsilon,\theta}}\right)
\end{equation}
\begin{eqnarray}
  |Z\mu_{ie},\theta,\epsilon \rangle & = & |Z\mu_{ie},\theta \rangle \nonumber \\
  & & - \sum_{j\not= i} \frac{\langle Z\mu_{je},\theta|\partial_\nu|Z\mu_{ie},\theta\rangle \tilde \epsilon^\nu(\theta)}{1-e^{\imath p_{\epsilon,\theta}(\chi_{je}-\chi_{ie})}} |Z\mu_{je},\theta \rangle \nonumber \\
  & & \quad + \mathcal O\left(\frac{\epsilon^2}{p_{\epsilon,\theta}}\right)
\end{eqnarray}
or conversely
\begin{equation}
  \chi_{ie} = \chi_{ie,\epsilon}(\theta) - \imath \langle Z\mu_i,\theta,\epsilon|\partial_\nu|Z\mu_i,\theta,\epsilon \rangle \frac{\tilde \epsilon^\nu(\theta)}{p_{\epsilon,\theta}} + \mathcal O\left(\frac{\epsilon^2}{p_{\epsilon,\theta}}\right)
\end{equation}
\begin{eqnarray}
  |Z\mu_{ie},\theta\rangle & = & |Z\mu_{ie},\theta,\epsilon\rangle \nonumber \\
  & & + \sum_{j\not= i} \frac{\langle Z\mu_{je},\theta,\epsilon|\partial_\nu|Z\mu_{ie},\theta,\epsilon \rangle \tilde \epsilon^\nu(\theta)}{1-e^{\imath p_{\epsilon,\theta}(\chi_{je,\epsilon}(\theta)-\chi_{ie,\epsilon}(\theta))}} |Z\mu_{je},\theta,\epsilon \rangle \nonumber \\
  & & \quad + \mathcal O(\epsilon^2)
\end{eqnarray}

\section{Physical meanings of the effective Hamiltonian}
\subsection{Approximate first recurrence Hamiltonian}\label{estimHeff}
In this section we want to exhibit concrete expressions for $H^{eff}_\epsilon(\theta)$. 

\subsubsection{Low frequency case}\label{lowfreq}
\begin{figure}
  \includegraphics[width=9cm]{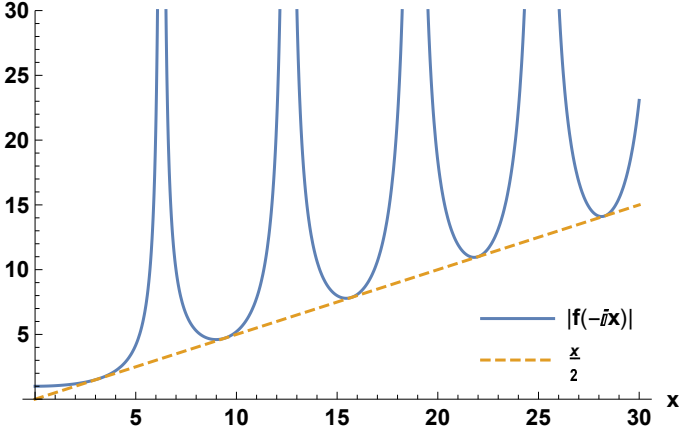}
  \caption{\label{fBCH} Plot of the function $x \mapsto \left|\frac{-\imath x}{1-e^{\imath x}} \right|$ appearing in the developments by the Baker-Campbell-Hausdorff formula (see \ref{BCH}).}
\end{figure}

Firstly we consider the low frequency regime where $\omega_0 \ll \omega_1$ (there are a lot of Rabi oscillations during a sampling period). This regime is consistent only with a kicked quantum system where the sampling period is the kick period. We have
\begin{eqnarray}
  U(\theta_p)...U(\theta_0) & = & e^{-\imath \frac{\omega_1}{\omega_0} \hat H} e^{-\imath V_p} ... e^{-\imath \frac{\omega_1}{\omega_0} \hat H} e^{-\imath V_0} \\
  & = & e^{-\imath (p+1)\frac{\omega_1}{\omega_0} \hat H} e^{-\imath \tilde V_p} ... e^{-\imath \tilde V_0}
\end{eqnarray}
with $V_n\equiv V(\theta_n)$ and $\tilde V_n \equiv e^{\imath n \frac{\omega_1}{\omega_0} \hat H} V_n e^{-\imath n \frac{\omega_1}{\omega_0} \hat H}$. $V(\theta)$ being supposed bounded, we have $\|V(\theta)\| \ll \frac{\omega_1}{\omega_0}$. It follows that
\begin{eqnarray}
  & & U(\theta_p)...U(\theta_0) \nonumber \\
  & & = e^{-\imath (p+1)\frac{\omega_1}{\omega_0} \hat H} e^{-\imath \sum_{n=0}^p \tilde V_n + \mathcal O\left(p\|V\|^2\right)} \\
  & & = e^{-\imath (p+1)\frac{\omega_1}{\omega_0} \hat H -\imath \sum_{n=0}^p f_{-\imath(p+1)\frac{\omega_1}{\omega_0} \hat H}[\tilde V_n]  + \mathcal O\left(p f(\imath p\frac{\omega_1}{\omega_0}\delta)\|V\|^2\right)}
\end{eqnarray}
where $f_X[Y] = f(\ad_X)[Y]$ with $f(x)=\frac{x}{1-e^{-x}}$ and $\ad_X[Y]=[X,Y]$, see \ref{BCH} concerning the Baker-Campbell-Hausdorff formula. $\delta$ is the gap between eigenvalues of $\hat H$ which maximizes $f(\imath p\frac{\omega_1}{\omega_0}\delta)$. By applying this result to the definition of $H^{eff}_\epsilon$ eq. (\ref{FRHeff}) we find
\begin{eqnarray}
  H^{eff}_\epsilon(\theta) & = & \frac{\omega_1}{\omega_0} \hat H \nonumber \\
  & & + \frac{1}{p_{\epsilon,\theta}} \sum_{n=0}^{p_{\epsilon,\theta}-1} f_{-\imath p_{\epsilon,\theta} \frac{\omega_1}{\omega_0} \hat H} [\tilde V_n(\theta)] \nonumber \\
  & & + \mathcal O\left(f(\imath p_{\epsilon,\theta}\frac{\omega_1}{\omega_0}\delta) \|V\|^2\right) \label{Hefflow}
\end{eqnarray}
where $\tilde V_n(\theta) = e^{\imath n \frac{\omega_1}{\omega_0} \hat H} V(\varphi^n(\theta)) e^{-\imath n \frac{\omega_1}{\omega_0} \hat H}$. 
To interprete this formula, we can re-express it in the eigenbasis of $\hat H$ ($\hat H|i\rangle = \lambda_i |i\rangle$):
\begin{eqnarray}
  H^{eff}_\epsilon(\theta) & = & \sum_i \left(\frac{\omega_1}{\omega_0} \lambda_i + \langle i|\bar V_\theta|i\rangle\right)|i\rangle\langle i| \nonumber \\
  & & + \sum_{i,j\not=i} f\left(-\imath p_{\epsilon,\theta} \frac{\omega_1}{\omega_0} (\lambda_i-\lambda_j)\right) \langle i|\bar V_\theta|j\rangle |i\rangle\langle j| \nonumber \\
    & & + \mathcal O\left(f(\imath p_{\epsilon,\theta} \frac{\omega_1}{\omega_0}\delta)\|V\|^2\right) \label{Hefflowrep}
\end{eqnarray}
where $\bar V_\theta = \frac{1}{p_{\epsilon,\theta}} \sum_{n=0}^{p_{\epsilon,\theta}-1} \tilde V_n(\theta)$ is the average of the interaction along $\Orb(\theta)$. We see that the effective Hamiltonian corresponds to the free Hamiltonian $\frac{\omega_1}{\omega_0} \hat H$ with its energies perturbed by the average interaction. This one induces also couplings between the free energy states which have magnitudes proportional to $|f\left(-\imath p_{\epsilon,\theta} \frac{\omega_1}{\omega_0} (\lambda_i-\lambda_j)\right)|$. The function $x \mapsto |f(-\imath x)|$ is plotted fig. \ref{fBCH}.
Since $p_{\epsilon,\theta} \frac{\omega_1}{\omega_0}$ is large, excepted if $\max_{i,j\not=i} |\lambda_i-\lambda_j|$ is realy very small, these couplings are strong. Moreover we have resonances if $p_{\epsilon,\theta} \frac{\omega_1}{\omega_0} |\lambda_i-\lambda_j| \in 2\pi \mathbb N^*$. Due to these resonances, the behavior of $H^{eff}_\epsilon(\theta)$ will be strongly sensitive to the value of $\frac{\omega_1}{\omega_0}$.

\subsubsection{High frequency case} \label{highfreq}
Now we consider the high frequency regime where $\omega_0 \gg \omega_1$ (there are a lot of samplings by the discrete description during one Rabi oscillation). This is the only one regime for a time discretization of a dynamics governed by a continuous time-dependent Hamiltonian where the sampling period is the discretization step. This regime can also be consistent with a kicked quantum system. We consider several subcases depending on the behaviour of $V(\theta)$.
\paragraph{Case 1: $V(\theta) \sim \mathcal O(\omega_1/\omega_0)$:}

We can compute $U(\theta_p)...U(\theta_0) = e^{-\imath \frac{\omega_1}{\omega_0} \hat H} e^{-\imath V_p} ... e^{-\imath \frac{\omega_1}{\omega_0} \hat H} e^{-\imath V_0}$ by using the Baker-Campbell-Hausdorff formula at the first order. We have then
\begin{equation} 
  H^{eff}_\epsilon(\theta) = \frac{\omega_1}{\omega_0}\hat H + \frac{1}{p_{\epsilon,\theta}} \sum_{n=0}^{p_{\epsilon,\theta}-1}  V(\varphi^n(\theta)) +  \mathcal O\left(\left(\frac{\omega_1}{\omega_0}\right)^2\right) \label{Heffhigh1}
\end{equation}

In that case the effective Hamiltonian is just the sum of free Hamiltonian and the average interaction.

\paragraph{Case 2: $[V(\theta),V(\theta')]=0$:}

We suppose that $[V(\theta_1),V(\theta_2)]=0$, $\forall \theta_1,\theta_2 \in \Orb(\theta)$. We have then by using the Baker-Campbell-Hausdorff formula explained \ref{BCH}:
\begin{eqnarray}
  & & U(\theta_p)...U(\theta_0) \nonumber \\
  & & =  e^{-\imath \frac{\omega_1}{\omega_0} \hat H} e^{-\imath V_p} ... e^{-\imath \frac{\omega_1}{\omega_0} \hat H} e^{-\imath V_0} \\
  & & =  e^{-\imath \sum_{n=0}^p V_n} e^{-\imath \frac{\omega_1}{\omega_0} \sum_{n=0}^p  \tilde H_n + \mathcal O \left(p\frac{\omega_1^2}{\omega_0^2}\right)} \\
  & & =  e^{-\imath \sum_{n=0}^p V_n -\imath \frac{\omega_1}{\omega_0} \sum_{n=0}^p f_{-\imath \sum_{q=0}^p V_q}[ \tilde H_n] + \mathcal O \left(p f(\imath p\delta) \frac{\omega_1^2}{\omega_0^2}\right)}
\end{eqnarray}
where $\tilde H_n = e^{\imath \sum_{q=0}^n V_q} \hat H e^{-\imath \sum_{q=0}^n V_q}$. $\delta$ is the gap between eigenvalues of the average interaction operator which maximizes $f(\imath p\delta)$. By applying this result to the definition of $H^{eff}_\epsilon$ eq. (\ref{FRHeff}) we find
\begin{eqnarray}
  H^{eff}_\epsilon(\theta) & = & \frac{1}{p_{\epsilon,\theta}} \sum_{n=0}^{p_{\epsilon,\theta}-1} V(\varphi^n(\theta)) \nonumber \\
  & & + \frac{\omega_1}{\omega_0} \frac{1}{p_{\epsilon,\theta}} \sum_{n=0}^{p_{\epsilon,\theta}-1} f_{-\imath \sum_{q=0}^{p_{\epsilon,\theta}-1} V(\varphi^q(\theta))} [ \tilde H_n(\theta)] \nonumber \\
  & & + \mathcal O\left(f(\imath p_{\epsilon,\theta}\delta)\left(\frac{\omega_1}{\omega_0}\right)^2\right) \label{Heffhigh2}
\end{eqnarray}
where $\tilde H_n(\theta) = e^{\imath \sum_{q=0}^n V(\varphi^q(\theta))} \hat H e^{-\imath \sum_{q=0}^n V(\varphi^q(\theta))}$.

\paragraph{Case 3: $V(\theta) = v(\theta)+W(\theta)$ with $[v(\theta),v(\theta')]=0$ and $W(\theta)\sim \mathcal O(\omega_1/\omega_0)$:}\label{case3}

This case is the combination of the two previous ones, with $V(\theta)=v(\theta)+W(\theta)$, where $W(\theta)\sim \mathcal O(\omega_1/\omega_0)$ and $[v(\theta_1),v(\theta_2)]=0$, $\forall \theta_1,\theta_2 \in \Orb(\theta)$. By using the Baker-Campbell-Hausdorff formula explained \ref{BCH}, we have
\begin{eqnarray}
  e^{-\imath \frac{\omega_1}{\omega_0} \hat H} e^{-\imath (v+W)} & = & e^{-\imath \frac{\omega_1}{\omega_0} \hat H} e^{-\imath f^{-1}_{-\imath v}[W]+ \mathcal O \left(\frac{\omega_1^2}{\omega_0^2}\right)} e^{-\imath v} \\
  & = & e^{-\imath \frac{\omega_1}{\omega_0} \hat H -\imath f^{-1}_{-\imath v}[W]+ \mathcal O \left(\frac{\omega_1^2}{\omega_0^2}\right)} e^{-\imath v}
\end{eqnarray}
We set $K = \frac{\omega_1}{\omega_0} \hat H + f^{-1}_{-\imath v}[W]$.
\begin{eqnarray}
  & & U(\theta_p)...U(\theta_0) \nonumber \\
  & & = e^{-\imath K_p+ \mathcal O \left(\frac{\omega_1^2}{\omega_0^2}\right)} e^{-\imath v_p} ... e^{-\imath K_0+ \mathcal O \left(\frac{\omega_1^2}{\omega_0^2}\right)} e^{-\imath v_0} \\
  & & = e^{-\imath \sum_{n=0}^p v_n} e^{-\imath \sum_{n=0}^p  \tilde K_n + \mathcal O \left(p\frac{\omega_1^2}{\omega_0^2}\right)} \\
  & & = e^{-\imath \sum_{n=0}^p v_n -\imath \sum_{n=0}^p f_{-\imath \sum_{q=0}^p v_q}[ \tilde K_n] + \mathcal O \left(pf(\imath p\delta)\frac{\omega_1^2}{\omega_0^2}\right)}
\end{eqnarray}
with $\tilde K_n = e^{\imath \sum_{q=0}^n v_q} K_n e^{-\imath \sum_{q=0}^n v_q}$. $\delta$ is the gap between eigenvalues of the average interaction operator which maximizes $f(\imath p\delta)$. By applying this result to the definition of $H^{eff}_\epsilon$ eq. (\ref{FRHeff}) we find
\begin{eqnarray}
  & & H^{eff}_\epsilon(\theta) \nonumber \\
  & = & \frac{1}{p_{\epsilon,\theta}}\sum_{n=0}^{p_{\epsilon,\theta}-1} v(\varphi^n(\theta)) \nonumber \\
  & & + \frac{1}{p_{\epsilon,\theta}}\sum_{n=0}^{p_{\epsilon,\theta}-1} f_{-\imath \sum_{q=0}^{p_{\epsilon,\theta}-1} v(\varphi^q(\theta))} [ \tilde K_n(\theta)] \nonumber \\
  & & + \mathcal O\left(f(\imath p_{\epsilon,\theta}\delta)\left(\frac{\omega_1}{\omega_0}\right)^2\right) \label{Heffhigh3}
\end{eqnarray}
with $\tilde K_n(\theta) = \frac{\omega_1}{\omega_0} e^{\imath \sum_{q=0}^n v(\varphi^q(\theta))} \hat H e^{-\imath \sum_{q=0}^n v(\varphi^q(\theta))}+f^{-1}_{-\imath v(\varphi^n(\theta))} \left[e^{\imath \sum_{q=0}^n v(\varphi^q(\theta))} W(\varphi^n(\theta)) e^{-\imath \sum_{q=0}^n v(\varphi^q(\theta))} \right]$.

\paragraph{Interpretation}

We consider here only the case 2, the case 1 being obvious and the case 3 being the superposition of the two first cases. Since we have a lot of sampling periods (a lot of short interactions) during a Rabi oscillation, the dynamics is dominated by the interaction operator. Let $\bar V_\theta =  \frac{1}{p_{\epsilon,\theta}} \sum_{n=0}^{p_{\epsilon,\theta}-1} V(\varphi^n(\theta))$ be the average of the interaction along $\Orb(\theta)$. Let $(|\bar a\rangle)_a$ be the eigenbasis of $\bar V_\theta$ ($\bar V_\theta |\bar a \rangle = \nu_a |\bar a \rangle$). Onto this basis, the effective Hamiltonian can be expressed as:
\begin{eqnarray}
  H^{eff}_\epsilon(\theta) & = & \sum_a (\nu_a+\frac{\omega_1}{\omega_0} \langle \bar a|\hat H|\bar a\rangle)|\bar a \rangle\langle \bar a| \nonumber \\
  & & + \frac{\omega_1}{\omega_0} \sum_{a,b\not=a} f(-ip_{\epsilon,\theta}(\nu_a-\nu_b)) \langle \bar a|\hat H|\bar b \rangle |\bar a\rangle \langle \bar b| \nonumber \\
  & & + \mathcal O\left(f(\imath p_{\epsilon,\theta}\delta)\left(\frac{\omega_1}{\omega_0}\right)^2\right) \label{Heffhighrep}
\end{eqnarray}
To understand this formula, it is instructive to consider the case where $V(\theta)$ is a kick operator of the form $V(\theta) = \lambda P(\theta)$ where $\lambda$ is the kick strengh and $P(\theta) = |w(\theta)\rangle\langle w(\theta)|$ is a rank-1 projection. This means that the interaction consists to kick the quantum system in the ``direction'' $|w(\theta) \rangle$. For example, with a two-level system, $|w(\theta)\rangle$ as a point onto the Bloch sphere defines a direction in the 3D-space. If the system is a spin kicked by ultra-short magnetic pulses, the direction defined onto the Bloch sphere is identified with the polarization direction of the magnetic field. If $\epsilon$ is sufficiently small, we have
\begin{equation}
  \bar V_\theta \simeq \lambda \int_{\Gamma_e} P(\theta) \frac{d\mu(\theta)}{\mu(\Gamma_e)} = \lambda \rho_e
\end{equation}
(with $\theta \in \Gamma_e$), where $\rho_e$ is a mixed state (a density matrix) corresponding to the average of $P(\theta)$ onto the ergodic component $\Gamma_e = \overline{\Orb(\theta)}$ endowed with the probability measure $\frac{d\mu(\theta)}{\mu(\Gamma_e)}$. We have then $\nu_a = \lambda p_a$ where $\{p_a\} = \Sp(\rho_e)$ are the probabilities to find the direction $|\bar a\rangle$ in the statistical mixture $\rho_e$. $\bar V_\theta$ can be then viewed as a kick of strenght $\lambda$ in a direction randomly chosen in $\{|\bar a \rangle\}_a$ with the probability law $\{p_a\}$. $\hat H$ induces perturbative corrections onto this probability law but it induces also quantum coherences of magnitudes proportional to $|f(-ip_{\epsilon,\theta}(\nu_a-\nu_b))|$, which are strong since $p_{\epsilon,\theta}$ is large. Anew, resonances occur if $p_{\epsilon,\theta} |\nu_a-\nu_b| \in 2\pi \mathbb N^*$. Note that in the case of a kick operator with kick delays as viewed in the introduction, $V(\theta)$ depends on $\frac{\omega_1}{\omega_0}$ (by relative phases in its representation on the eigenbasis of $\hat H$). The behavior of $H^{eff}_\epsilon(\theta)$ will be also in this case strongly sensitive to the value of $\frac{\omega_1}{\omega_0}$ because of these resonances.

\subsubsection{About the accuracy of the approximations}
The formulae for $H_\epsilon^{eff}(\theta)$ found in this section are very rough approximations. We can only consider them for qualitative discussions or for physical interpertations. We cannot use them in qualitative discussions, especially for numerical computations. The reason of this bad accuracy are the error magnitude of the order of $f(\imath p_{\epsilon,\theta}\delta) \left(\frac{\omega_1}{\omega_0}\right)^2$. This one is reasonable only if the almost-period $p_{\epsilon,\theta}$ is small (and out of the resonances associated with the poles of $f$). But, with an almost-periodic or a chaotic dynamics, it is small only if $\epsilon$ is large. In that case, it is the almost-periodicity assumption which is very rough. We have a small $p_{\epsilon,\theta}$ only for strictly periodic dynamics or for dynamics extremely close to a periodic dynamics.\\
To make numerical computations of the effective Hamiltonians, it is more usefull to use directly the definition of the first recurrence Hamiltonian eq. \ref{FRHeff} or to solve eq. \ref{SKquasistate} for example by the method explained \ref{HKcomp}.

\subsection{Expected behaviours with almost-periodically driven systems} \label{expected}
An orbit $\Orb(\theta)$ is characterized by three quantities. The first one is its almost-period $p_{\epsilon,\theta}$. The second one is its mean diameter $\angle \Orb(\theta)$, the characteristic distance onto $\Gamma$ between two opposite points of $\Orb(\theta)$. It measures the mean dispersion of $\Orb(\theta)$ onto $\Gamma$. And finally $\underline \lambda_\theta$ the Lyapunov exponent (\cite{Haake}) which measures the ``chaoticity'' of the dynamics starting in the neighbourhood of $\theta$. In this section we want to present expected dynamical behaviors of almot-periodically driven quantum systems with respect to the values of these quantities, when the interaction $V(\theta)$ is perturbative.\\

By eq. \ref{FRHeff} the dynamics generated by $H^{eff}_\epsilon(\theta)$ during $p_{\epsilon,\theta}$ steps is the same than $U(\varphi^{p_{\epsilon,\theta}-1}(\theta))...U(\theta)$ and then $\forall \psi$ ($\|\psi\|=1$)
\begin{eqnarray}
  F_1^{str}(\theta) & = & |\langle \psi|e^{\imath (p_{\epsilon,\theta}+1)H^{eff}_{\epsilon}(\theta)} \mathbf{U}_{p_{\epsilon,\theta}}(\theta)|\psi\rangle|^2 \\
  & = & 1 + \mathcal O(\epsilon)
\end{eqnarray}
where $\mathbf{U}_n(\theta)= U(\varphi^n(\theta))...U(\theta)$. For a $p$-cyclic orbit, $H^{eff}(\theta)$ governs the global regime (on the time scale of $p$ steps) where the transient regime (with time scale lower than $p$ steps) is erased. The stroboscopic dynamics governed by $H^{eff}(\theta)$ is the exact true dynamics of the quantum system with the stroboscopic period $p$. For almost-periodic orbit, we can then be interested by the stroboscopic fidelity of the dynamics governed by $H^{eff}_\epsilon(\theta)$:
\begin{equation} \label{strobFid}
  F_n^{str}(\theta) = |\langle \psi|e^{\imath (np_{\epsilon,\theta}+1)H^{eff}_{\epsilon}(\theta)} \mathbf{U}_{np_{\epsilon,\theta}}(\theta)|\psi\rangle|^2
\end{equation}
As previously said, we have exactly $F_n^{str}(\theta) = 1$ ($\forall n \in \mathbb N$) for a $p$-cyclic orbit. Under what conditions do we have $F_n^{str}(\theta) \simeq 1$ with an almost-periodic orbit? Firstly, this needs that $\underline \lambda(\theta) = 0$, because if the flow is chaotic, due to the sensitivity to initial conditions (\cite{Haake}) $\Orb(\varphi^{p_{\epsilon,\theta}}(\theta))$ exponentially separates from $\Orb(\theta)$ as $e^{n \underline \lambda_\theta} \epsilon$ (even if $\|\varphi^{p_{\epsilon,\theta}}(\theta)- \theta\|< \epsilon$). In particular, we have $p_{\epsilon,\varphi^{p_{\epsilon,\theta}}(\theta)} \not= p_{\epsilon,\theta}$: the recurrence of the flow in the neighbourhood of $\theta$ is erratic, the almost-period drastically change at each recurrence. We can think that $F_n^{str}(\theta)$ decreases if $\frac{\omega_1}{\omega_0}$ increases (or equivalently the approximation $F_n^{str}(\theta) \simeq 1$ is valid until a smaller value of $n$ if $\frac{\omega_1}{\omega_0}$ is larger). Indeed, as we can see it in the low frequency regime ($\frac{\omega_1}{\omega_0} \gg 1$, eq. \ref{Hefflowrep}), the large factors $|f(-\imath p_{\epsilon,\theta} \frac{\omega_1}{\omega_0} (\lambda_i-\lambda_j))|$ reinforce the $\theta$-dependent couplings $\langle i|\bar V_\theta|j \rangle$ which become non-perturbative. So, the small difference between $\langle i|\bar V_\theta|j \rangle$ and $\langle i|\bar V_{\varphi^{p_{\epsilon,\theta}}(\theta)}|j \rangle$ will be amplified by these factors (especially close to the resonances). This problem does not occur in the high frequency regime. Finally we can think that the approximation $F_n^{str}(\theta) \simeq 1$ with small values of $\frac{\omega_1}{\omega_0}$ is better with not too large diameters $\angle \Orb(\theta)$. In the high frequency regime, $\bar V_\theta - \bar V_{\varphi^{n p_{\epsilon,\theta}}(\theta)} \sim \mathcal O(\mu(S_n))$ where $S_n \subset \Gamma$ is the region delimited by $\overline{\Orb(\theta)}$ and $\overline{\Orb(\varphi^{n p_{\epsilon,\theta}}(\theta))}$. It is more probable that $\mu(S_n)$ quickly becomes large with $n$ if $\angle \Orb(\theta)$ is large.\\

By eq. \ref{FRHeff} and \ref{SKquasistate} we have
\begin{equation}
  |\langle Z\mu_{ie},\varphi^{n+1}(\theta)|\mathbf{U}_n(\theta)|Z\mu_{ie},\theta\rangle|^2=1
\end{equation}
(with $\theta \in \Gamma_e$). $\mathbf{U}_n(\theta)$ describes the complete dynamics whereas $e^{-\imath n H^{eff}_\epsilon(\theta)}$ describes the global dynamics without the transient fluctuations occuring at time scale lower than the almost-period. As eigenvector of $e^{-\imath H^{eff}(\theta)}$, the quasi-energy state $|Z\mu_{ie},\theta\rangle$ is then the steady state of the global dynamics. Its evolution could be almost steady with the fluctuations associated with the transient regime. Let the survival probability of the quasi-energy state be:
\begin{equation}
  P^{surv}_n(\theta) = |\langle Z\mu_{ie},\theta|\mathbf{U}_n(\theta)|Z\mu_{ie},\theta \rangle|^2
\end{equation}
We have $P^{surv}_{p_{\epsilon,\theta}}(\theta) = 1+\mathcal O(\epsilon)$ and $P^{surv}_{np_{\epsilon,\theta}}(\theta) \simeq 1$ in the same conditions that the previous discussion (with the choice $|\psi\rangle = |Z\mu_{ie},\theta \rangle$). The quasi-energy states are then states of the quantum system which are almost recurrent. They are then very important as the cyclic quantum states associated with the Floquet theory (which they are a generalization) and are associated with some quantum phenomena as the quantum revivals \cite{Bocchieri,Eberly}. But moreover we can hope that $P^{surv}_n(\theta) \approx 1$ if the fluctuations generated by $V(\varphi^n(\theta)) - \bar V_\theta$ on $|Z\mu_{ie},\theta\rangle$ are small ($\bar V_\theta$ being the average interaction operator along $\Orb(\theta)$). This needs that $\angle \Orb(\theta) \ll 1$, because the variations of $V(\varphi^n(\theta))$ will be large if the orbit is large. We can think that the assumption $P^{surv}_n(\theta) \approx 1$ is easier satisfied if $p_{\epsilon,\theta}$ is not large, in order to the duration of the transient regime be short. Moreover, if $V(\theta)$ depends on $\frac{\omega_1}{\omega_0}$ as for a kick operator with kick delay (as explained in the introduction), we must have $\frac{\omega_1}{\omega_0} \not\gg 1$ otherwise the presence of fast oscillating phases in $V(\varphi^n(\theta))$ induces strong difference between $V(\varphi^n(\theta))$ and $\bar V_\theta$. And finally, to have $P^{surv}_n(\theta) \approx 1$ with $n>p_{\epsilon,\theta}$, we need $\underline \lambda_\theta = 0$ because of the sensitivity to initial conditions of the chaotic flows which implies that $\bar V_{\varphi^{p_{\epsilon,\theta}}(\theta)} \not\simeq \bar V_{\theta}$. 

\section{Illustration}
In order to illustrate the concepts presented in this paper, we consider the following driven quantum system: a two-level quantum system defined by the canonical basis $(|0\rangle,|1\rangle)$ and the free Hamiltonian $H_0 = \hbar\omega_1|1\rangle\langle 1|$ (for example a $\frac{1}{2}$-spin system with Zeeman effect), kicked with a frequency $\omega_0$ following the kick operator $V(\theta)=\lambda|w(\theta^1,\theta^2)\rangle\langle w(\theta^1,\theta^2)|$ with $|w(\theta^1,\theta^2)\rangle = \cos\theta^1|0\rangle + e^{\imath \frac{\omega_1}{\omega_0} \theta^2}\sin\theta^1 |1 \rangle$. $\lambda=0.1$ is the dimensionless kick strenght, $\theta^2$ is the angular kick delay, and $(\theta^1,\frac{\omega_1}{\omega_0} \theta^2)$ defines the kick direction in a spherical coordinates with the $z$-axis corresponding to the direction of the Zeeman magnetic field. We have then $U(\theta) = e^{-\imath \frac{\omega_1}{\omega_0} \hat H} e^{-\imath V(\theta)}$ (with $\hat H = 2\pi|1\rangle\langle 1|$). Since $V$ is proportional to a rank 1 projection, it is easy to compute the matrix exponentials: we have in the basis $(|0\rangle,|1\rangle)$:
\begin{eqnarray}
  & &e^{-\imath \frac{\omega_1}{\omega_0} \hat H} = \left(\begin{array}{cc} 1 & 0 \\ 0 & e^{-\imath 2 \pi \frac{\omega_1}{\omega_0}} \end{array} \right) \\
  & & e^{-\imath V(\theta)} =  1_2 +(e^{-\imath \lambda}-1) \times \nonumber \\
  & & \left(\begin{array}{cc} \cos^2 \theta^1 & \frac{e^{\imath \frac{\omega_1}{\omega_0} \theta^2}}{2} \sin(2\theta^1) \\  \frac{e^{-\imath \frac{\omega_1}{\omega_0} \theta^2}}{2} \sin(2\theta^1) & \sin^2 \theta^1 \end{array} \right)
\end{eqnarray}
Remark : $V(\theta) = \lambda e^{\imath \frac{H_0}{\hbar \omega_0} \theta^2} |w(\theta^1,0)\rangle\langle w(\theta^1,0)| e^{-\imath \frac{H_0}{\hbar \omega_0} \theta^2} = \lambda |w(\theta^1,0)\rangle\langle w(\theta^1,0)| + \imath \frac{\lambda \theta^2}{\hbar \omega_0} [H_0, |w(\theta^1,0)\rangle\langle w(\theta^1,0)|] + \mathcal O(\omega_1^2/\omega_0^2)$. In the high frequency regime with constant $\theta^1$ the system belongs to the case 3 viewed in section \ref{case3}.

\begin{figure}
  \includegraphics[width=9cm]{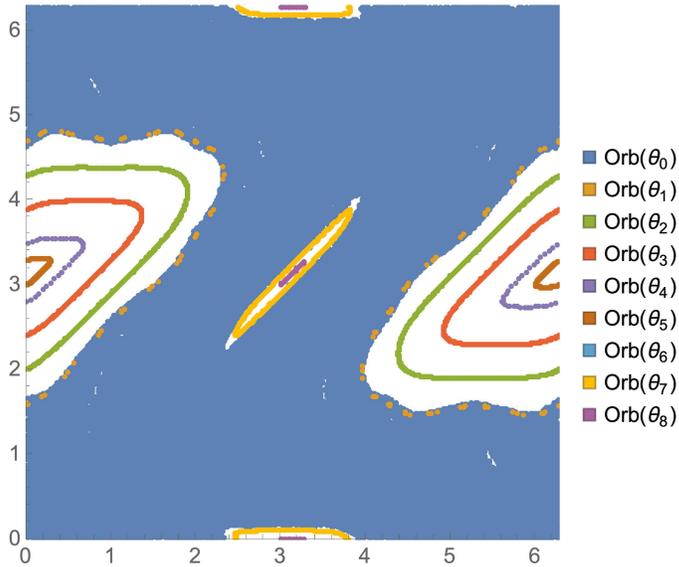}
  \caption{\label{StdMap} Phase portrait of the Chirikov standard map onto the torus $\mathbb T^2$, with 9 orbits considered in the simulations.}
\end{figure}

The phase space is $\Gamma=\mathbb T^2$ (the 2-torus generated by $\theta=(\theta^1,\theta^2) \in [0,2\pi]^2$). We consider the uniform measure onto $\mathbb T^2$: $d\mu(\theta) = \frac{d\theta^1d\theta^2}{(2\pi)^2}$ (with $\mathscr T$ the Borelian $\sigma$-algebra). The flow $\varphi \in \mathrm{Aut} (\mathbb T^2)$ is then an invariant automorphism of the 2-torus. We choose the Chirikov standard map defined by:
\begin{equation}
  \varphi(\theta) = \left(\begin{array}{lc} \theta^1+K \sin(\theta^2) & \mod 2\pi \\ \theta^1+\theta^2+K\sin(\theta^2) & \mod 2\pi \end{array} \right)
\end{equation}
with $K=2$. Historically the standard map models the behaviour of a kicked rotator, but it can be considered as an universal model for kicked nonlinear oscillators \cite{Chirikov}. The model used in this illustration can correspond to a spin system submitted to ultrafast magnetic pulses produced by a device modeled by a nonlinear oscillator. The device receives a periodic control signal assimilated to regular kicks of frequency $\omega_0$. But the device response is not instantaneous and there is a delay $\theta^2$ before the pulse emission. Due to the nonlinear character of the device oscillations, this delay changes at each cycle. Moreover, the pulses are emitted with a polarization angle $\theta^1$ which also changes at each cycle due to the nonlinearity. So in place of a regular pulse train of frequency $\omega_0$, the pulse train is only almost-periodic with an almost-period $p_{\epsilon,\theta_0}$ depending of the initial condition of the device. The possible goal of a study of this system with the SK effective Hamiltonian is to understand the effect of the nonlinearity at long-term onto the kicked spin.\\
The phase portrait of the standard map flow is plotted fig. \ref{StdMap}.
We can distinguish three different areas. The first one is the chaotic sea (in blue fig. \ref{StdMap}) which is an ergodic component $\Gamma_0 = \overline{\Orb(\theta_0)}$ associated with a chaotic orbit. The fixed point $(0,0)$ is embedded in this chaotic sea. A big island of stability centered on the fixed point $(0,\pi)$ is constituted by quasi-periodic orbits. The irrationally related frequencies of these orbits are numerous. We consider five ergodic components $\{\Gamma_e = \overline{\Orb(\theta_e)}\}_{e=1,...,5}$ in this island for the numerical study. And finally, we have a double small island of stability with two connected components centered on $(\pi,0)$ and $(\pi,\pi)$ ($(\pi,0) \leftrightarrows (\pi,\pi)$ is a 2-cyclic orbit). We consider these two components as part of a same island because the inner orbit jump from a component to the other one. We consider three ergodic components $\{\Gamma_e = \overline{\Orb(\theta_e)}\}_{e=6,7,8}$ in this island. We have choosen an initial point $\theta_e$ in each ergodic component to start the dynamics. Except for $\Gamma_0$, this choice has no influence onto the results. The properties of the nine considered orbits are reported table \ref{orbits}.
\begin{table}
  \caption{\label{orbits} Properties of the orbits used in the dynamics, with $\epsilon = 10^{-2}$. The almost-period $p_{\epsilon,\theta_e}$ does not depend on the choice of $\theta_e \in \Gamma_e$ except for $\Gamma_0$. In some simulations we consider also $\epsilon = 10^{-1}$ for $\Gamma_0$, in that case $p_{\epsilon,\theta_0} = 734$. The mean diameter $\angle \Orb(\theta_e)$ has been estimated as the average between the maximum and the minimum diameters of the almost closed orbits in the islands of stability. Since the chaotic sea covers a large part of $\mathbb T^2$ its mean diameter is $2\pi$.}
  \begin{tabular}{|c|r|r|r|l|}
      \hline
      e & $p_{\epsilon,\theta_e}$ & $\angle \Orb(\theta_e)$ & $\underline \lambda_{\theta_e}$ & region \\
      \hline
      0 & 25801 & $2\pi$ & $0.415$ & chaotic sea \\
      1 & 108 & $4.7$ & $0$ & big island border \\
      2 & 926 & $3.8$ & $0$ & big island \\
      3 & 845 & $2.6$ & $0$ & big island \\
      4 & 69 & $1.2$ & $0$ & big island \\
      5 & 385 & $0.96$ & $0$ & big island center \\
      6 & 26 & $2$ & $0$ & double small island border \\
      7 & 430 & $2$ & $0$ & double small island \\
      8 & 42 & $0.5$ & $0$ & double small island center\\
      \hline
  \end{tabular}
\end{table}

In the numerical simulations, we can compute $U^{eff}_\epsilon(\theta) = e^{-\imath H^{eff}_\epsilon(\theta)}$ by two manners. The first one consists to use eq. \ref{FRHeff}:
\begin{equation}
  U^{eff}_{\epsilon}(\theta_e) = \sqrt[p_{\epsilon,\theta_e}]{U(\varphi^{p_{\epsilon,\theta_e}-1}(\theta_e))...U(\theta_e)}
\end{equation}
The second one consists to use the method presented in \ref{HKcomp}. The two ones provide very similar numerical results.

\subsection{Stroboscopic fidelity of the dynamics}
\begin{figure}
  \includegraphics[width=9cm]{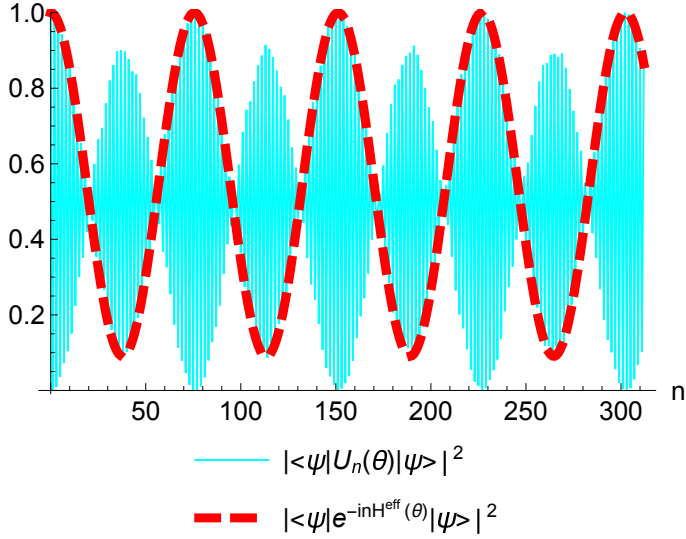}
  \caption{\label{compDyn} Comparision of the true dynamics $|\langle \psi|\mathbf{U}_n(\theta_6)|\psi \rangle|^2$ and the effective dynamics $|\langle \psi|e^{-\imath n H^{eff}_\epsilon(\theta_6)}|\psi \rangle|^2$ for the orbit $e=6$ during 12 almost-periods, with $\frac{\omega_1}{\omega_0} = 3.5$ and $|\psi\rangle = \frac{1}{\sqrt 2}(|0\rangle+|1\rangle)$.}
\end{figure}

As previously explained, the dynamics governed by $H^{eff}_\epsilon(\theta)$ is the global dynamics without the fluctuations of the transient regime with small time scale. Fig. \ref{compDyn} gives an illutration of this.
In this example, the global dynamics is the envelope of the complete dynamics, but it is not always the case (it is necessary to tune $\frac{\omega_1}{\omega_0}$ to have this simple behaviour). As viewed section \ref{estimHeff}, the average interaction $\bar V_{\theta_6} = \frac{1}{p_{\theta_6,\epsilon}} \sum_{n=0}^{p_{\theta_6,\epsilon}} V(\varphi^n(\theta_6))$ plays an important role in the effective dynamics. Here $\bar V_{\theta_6} \simeq \lambda \oint_{\Gamma_6} |w\rangle \langle w|\frac{d\ell}{\ell(\Gamma_6)} = \lambda \rho_6$ ($\Gamma_6$ being assimilated to a closed path of length $\ell(\Gamma_6)$) where $\rho_6 = p_0|0\rangle \langle 0| + p_1|1\rangle \langle 1|$ is the density matrix representing the fact that the evolution along $\Gamma_6$ can in part be viewed as a kicking of the spin in up ($|0\rangle\langle  0|$) or down ($|1\rangle\langle 1$) directions (i.e. magnetic pulse polarizations) randomly choosen with the probability law $\{p_0,p_1\}$. The Rabbi oscillations of the spin system induced by the free Hamiltonian $\frac{\omega_1}{\omega_0} \hat H$, which are of frequency $2\pi \frac{\omega_1}{\omega_0}$, are the fast oscillations of the carrier wave in fig. \ref{compDyn}. As a transcient behaviour, it is erased in the effective dynamics. By erasing these oscillations, the effective dynamics focus on the behaviour related to $\bar V_{\theta_6}$ permitting the study of the long-term effects of the nonlinearity in the device. Roughly speacking, the effective dynamics corresponds then to randomly kicked spin following the law defined by $\rho_6$, and then $e^{-\imath n H^{eff}_\epsilon(\theta_6)}|\psi\rangle \approx \frac{1}{\sqrt 2}(e^{-\imath n p_0 \lambda}|0\rangle +  e^{-\imath n p_1 \lambda}|1\rangle)$ (since $p_0$ is the ratio of kicks in the up direction). The survival probability is then $|\langle \psi|e^{-\imath n H^{eff}_\epsilon(\theta_6)}|\psi\rangle|^2 \approx \frac{1+\cos((p_0-p_1)n\lambda)}{2}$ and oscillates with a period $\frac{2\pi}{(p_0-p_1)\lambda}$. If all kicks had been in the up direction, the survival probability oscillation period would have been $\frac{2\pi}{\lambda} \simeq 63$. We see fig. \ref{compDyn}, than due the kicks in the opposite direction the period is slightly larger ($p_0$ is indeed close to 1, since $\Gamma_6$ is closed to the the 2-cyclic orbit $(\pi,0) \leftrightarrows (\pi,\pi)$ for which $V(\pi,0)=V(\pi,\pi)=\lambda|0\rangle \langle 0|$).

\begin{figure}
  \includegraphics[width=9cm]{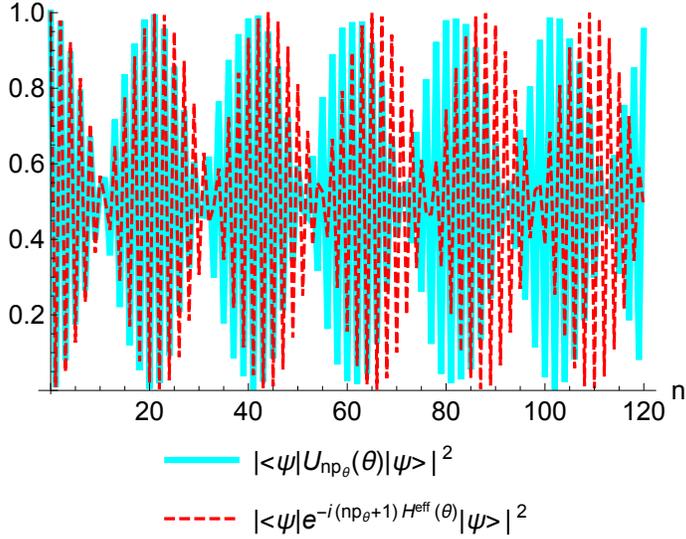}
  \caption{\label{compStrob} Comparision of the true stroboscopic survival probability $|\langle \psi|\mathbf{U}_{np_{\epsilon,\theta_5}}(\theta_5)|\psi \rangle|^2$ and the effective stroboscopic survival probability $|\langle \psi|e^{-\imath (np_{\epsilon,\theta_5}+1) H^{eff}_\epsilon(\theta_5)}|\psi \rangle|^2$ for the orbit $e=5$, with $\frac{\omega_1}{\omega_0} = 4.5$ and $|\psi\rangle = \frac{1}{\sqrt 2}(|0\rangle+|1\rangle)$. We see a small dephasing occuring with a large number of almost-periods. The average stroboscopic fidelity during 12 almost-periods is $99.7\%$ whereas due to the dephasing, during 120 almost-periods, it is only $93.9\%$. This dephasing is due to the fact that numerically we work with the first recurrence Hamiltonian $H^{eff}_\epsilon$ which is only an approximation of the SK Hamiltonian up to an error $\epsilon/p_\epsilon$ (eq. \ref{relatHeff}). So after $n$ almost-periods, the error in the evolution operator is of magnitude $n\epsilon$. Because $\epsilon = 10^{-2}$, this error induced the dephasing for large values of $n$.}
\end{figure}

It is more interesting to study the stroboscopic fidelity eq. \ref{strobFid}. As illustration, we can see fig. \ref{compStrob}.
In order to enlighten the efficiency of the effective description for the stroboscopic dynamics we have compute the avergage stroboscopic fidelity:
\begin{equation}
  \overline{F^{str}(\theta)} = \frac{1}{N+1} \sum_{n=0}^N F_n^{str}(\theta)
\end{equation}
at short term ($N=12$ almost-periods) and at long term ($N=120$ almost-periods), for the three regimes (low, medium and high frequency). Since we have shown that the behaviours are very sensitive to the value of $\frac{\omega_1}{\omega_0}$ we have considered for each regime three different values of the period ratio. The results are presented tables \ref{strobFid12} and \ref{strobFid120}.
\begin{table}
  \caption{\label{strobFid12} Avergage stroboscopic fidelity of the effective dynamics during 12 almost-periods with $|\psi\rangle = \frac{1}{\sqrt 2}(|0\rangle+|1\rangle)$, for the different orbits and for different values of $\frac{\omega_1}{\omega_0}$. We writte in bold the good results ($\geq 97\%$), in sans serif style the correct results ($\geq 90\%$), in normal style the middling results ($\geq 75\%$) and in  italic the bad results ($<75\%$).}
  \scriptsize
  \begin{tabular}{|c|c|c|c|c|c|c|c|c|c|}
    \hline
    $e$ & \multicolumn{3}{|c|}{High freq. $\frac{\omega_1}{\omega_0}\ll 1$} & \multicolumn{3}{|c|}{Medium freq. $\frac{\omega_1}{\omega_0} \sim 1$} & \multicolumn{3}{|c|}{Low freq. $\frac{\omega_1}{\omega_0}\gg 1$} \\
    \cline{2-10}
     & $\frac{\sqrt{2}}{100}$ & $0.03$ & $0.04$ & $\sqrt{2}$ & $3.4$ & $4.5$ & $100\sqrt{2}$ & $101.3$ & $104.5$ \\
    \hline
    0 & $\mathit{74.3\%}$ & $\mathit{74.4\%}$ & $79.1\%$ & $77.6\%$ & $\mathit{74.2\%}$ & $75.0\%$ & $\mathit{74.6\%}$ & $78.3\%$ & $\mathit{61.6\%}$ \\
    \hline
    1 & $\mathbf{100\%}$ & $\mathbf{100\%}$ & $\mathbf{98.0\%}$ & $\mathbf{100\%}$ & $\mathbf{99.6\%}$ & $\mathbf{99.9\%}$ & $\mathbf{98.1\%}$ & $83.7\%$ & $\mathbf{98.0\%}$ \\
    \hline
    2 & $\mathbf{100\%}$ & $\mathbf{100\%}$ & $\mathbf{100\%}$ & $\mathbf{98.8\%}$ & $\mathbf{98.6\%}$ & $\mathsf{91.8\%}$ & $\mathit{69.4\%}$ & $\mathit{74.8\%}$ & $\mathit{73.2\%}$ \\
    \hline
    3 & $\mathbf{100\%}$ & $\mathbf{100\%}$ & $\mathbf{100\%}$ & $\mathbf{99.1\%}$ & $\mathbf{98.7\%}$ & $\mathbf{98.6\%}$ & $\mathit{73.7\%}$ & $\mathbf{99.4\%}$ & $\mathsf{95.8\%}$ \\
    \hline
    4 & $\mathbf{100\%}$ & $\mathbf{100\%}$ & $\mathbf{100\%}$ & $\mathbf{100\%}$ & $\mathbf{100\%}$ & $\mathbf{99.7\%}$ & $\mathbf{99.4\%}$ & $\mathbf{99.9\%}$ & $\mathbf{99.9\%}$ \\
    \hline
    5 & $\mathbf{100\%}$ & $\mathbf{100\%}$ & $\mathbf{100\%}$ & $\mathbf{100\%}$ & $\mathbf{99.8\%}$ & $\mathbf{99.7\%}$ & $\mathbf{99.9\%}$ & $\mathbf{99.8\%}$ & $\mathit{70.8\%}$ \\
    \hline
    6 & $\mathbf{100\%}$ & $\mathbf{100\%}$ & $\mathbf{100\%}$ & $\mathbf{100\%}$ & $\mathbf{99.8\%}$ & $\mathbf{100\%}$ & $\mathbf{98.9\%}$ & $\mathbf{98.9\%}$ & $\mathbf{97.9\%}$ \\
    \hline
    7 & $\mathbf{100\%}$ & $\mathbf{100\%}$ & $\mathbf{100\%}$ & $\mathbf{100\%}$ & $\mathbf{100\%}$ & $\mathbf{100\%}$ & $\mathsf{92.9\%}$ & $\mathbf{99.8\%}$ & $\mathbf{99.7\%}$ \\
    \hline
    8 & $\mathbf{100\%}$ & $\mathbf{100\%}$ & $\mathbf{100\%}$ & $\mathbf{100\%}$ & $\mathbf{100\%}$ & $\mathbf{100\%}$ & $\mathsf{99.6\%}$ & $\mathbf{99.7\%}$ & $\mathbf{99.9\%}$ \\
    \hline
  \end{tabular}
\end{table}
\begin{table}
  \caption{\label{strobFid120} Same as table \ref{strobFid12} but with averaging during 120 almost-periods. Moreover for the orbit $e=0$ the presented results correspond to $\epsilon=10^{-1}$ (whereas $\epsilon=10^{-2}$ for the other orbits).}
  \scriptsize
  \begin{tabular}{|c|c|c|c|c|c|c|c|c|c|}
    \hline
    $e$ & \multicolumn{3}{|c|}{High freq. $\frac{\omega_1}{\omega_0}\ll 1$} & \multicolumn{3}{|c|}{Medium freq. $\frac{\omega_1}{\omega_0} \sim 1$} & \multicolumn{3}{|c|}{Low freq. $\frac{\omega_1}{\omega_0}\gg 1$} \\
    \cline{2-10}
     & $\frac{\sqrt{2}}{100}$ & $0.03$ & $0.04$ & $\sqrt{2}$ & $3.4$ & $4.5$ & $100\sqrt{2}$ & $101.3$ & $104.5$ \\
    \hline
    0 & $\mathit{61.5\%}$ & $\mathit{64.1\%}$ & $\mathit{68.1\%}$ & $\mathit{68.1\%}$ & $\mathit{69.2\%}$ & $\mathit{66.4\%}$ & $\mathit{72.0\%}$ & $\mathit{66.5\%}$ & $\mathit{72.5\%}$ \\
    \hline
    1 & $\mathbf{100\%}$ & $\mathbf{100\%}$ & $\mathit{61.0\%}$ & $\mathbf{99.5\%}$ & $\mathbf{98.0\%}$ & $\mathbf{99.9\%}$ & $\mathsf{91.4\%}$ & $\mathit{65.9\%}$ & $\mathit{62.5\%}$ \\
    \hline
    2 & $\mathbf{99.9\%}$ & $\mathbf{99.9\%}$ & $\mathbf{99.9\%}$ & $79.7\%$ & $\mathit{71.6\%}$ & $\mathit{64.1\%}$ & $\mathit{72.6\%}$ & $\mathit{64.2\%}$ & $\mathit{69.8\%}$ \\
    \hline
    3 & $\mathbf{99.9\%}$ & $\mathbf{99.8\%}$ & $\mathbf{99.7\%}$ & $88.7\%$ & $\mathit{60.0\%}$ & $\mathsf{91.0\%}$ & $\mathit{70.2\%}$ & $\mathsf{93.3\%}$ & $80.1\%$ \\
    \hline
    4 & $\mathbf{100\%}$ & $\mathbf{100\%}$ & $\mathbf{100\%}$ & $\mathbf{99.2\%}$ & $\mathbf{98.9\%}$ & $76.6\%$ & $\mathsf{93.7\%}$ & $\mathsf{96.6\%}$ & $\mathbf{97.3\%}$ \\
    \hline
    5 & $\mathbf{97.9\%}$ & $\mathbf{98.4\%}$ & $\mathbf{98.4\%}$ & $\mathbf{98.9\%}$ & $\mathsf{96.2\%}$ & $\mathsf{93.9\%}$ & $\mathbf{92.6\%}$ & $\mathbf{99.1\%}$ & $\mathit{71.1\%}$ \\
    \hline
    6 & $\mathbf{99.8\%}$ & $\mathbf{99.8\%}$ & $\mathbf{99.8\%}$ & $\mathbf{99.6\%}$ & $\mathsf{95.2\%}$ & $\mathbf{99.8\%}$ & $\mathit{60.6\%}$ & $\mathsf{91.4\%}$ & $\mathit{66.6\%}$ \\
    \hline
    7 & $\mathbf{100\%}$ & $\mathbf{100\%}$ & $\mathbf{100\%}$ & $\mathbf{100\%}$ & $\mathbf{99.5\%}$ & $\mathbf{100\%}$ & $\mathit{61.6\%}$ & $76.3\%$ & $\mathsf{97.1\%}$ \\
    \hline
    8 & $\mathbf{99.3\%}$ & $\mathbf{99.3\%}$ & $\mathbf{99.3\%}$ & $\mathbf{99.3\%}$ & $\mathbf{99.3\%}$ & $\mathbf{99.2\%}$ & $\mathbf{98.4\%}$ & $\mathsf{94.7\%}$ & $\mathbf{99.0\%}$ \\
    \hline
  \end{tabular}
\end{table}
These results are in accordance with the discussion of the section \ref{expected}. Except for the case of the chaotic orbit $\Orb(\theta_0)$, the average stroboscopic fidelity is large for the high frequency regime whereas it is good in the low frequency regime only at short term. This situation where the effective Hamiltonian describes very correctly the dynamics at high frequency but not at low frequency exists also for periodic systems, as in the system described in ref. \cite{Poletti}. In this one, the bad behaviour at low frequency is interpreted as the result of resonances between the quantum system and the periodic control. We can propose a similar interpretation here, as viewed section \ref{lowfreq}, at low frequency, resonances between the quantum system and the almost periodic control occur if $\frac{\omega_1}{\omega_0} \in \frac{\mathbb N^*}{p_{\epsilon,\theta}}$ (fig. \ref{fBCH}). Due to the closeness of these resonances, the effective Hamiltonian (depending only on fundamental quasi-energy states) fails to describe completely the dynamics, and an approach taking into account all quasi-energy states (as in ref. \cite{Viennot3}) is needed. In contrast, at high frequency, the possible resonances evoked section \ref{highfreq}, $\lambda|p_0-p_1| \in \frac{2\pi \mathbb N^*}{p_{\epsilon,\theta}}$,  depend only on the characteristics of the interaction $\bar V_\theta$ (and not on the quantum system itself) and are unlikely for a generic orbit.\\
Moreover the results seem better in the double small island and in the center of the big island, confirming that with a too large orbit diameter $\angle \Orb(\theta)$ the stroboscopic fidelity is lower.  

\subsection{Almost steady states}
\begin{figure}
  \includegraphics[width=9cm]{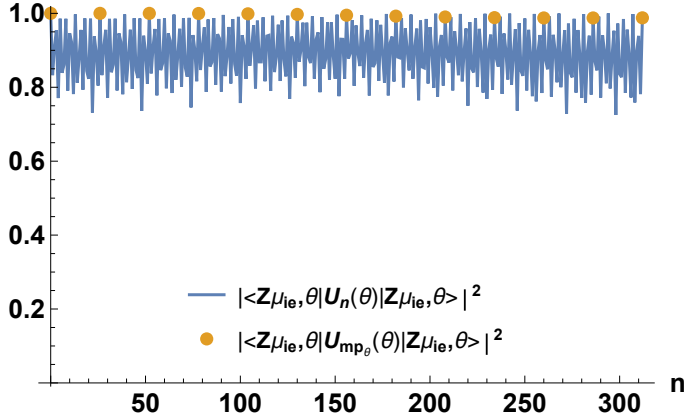}
  \caption{\label{survprobZmu} Survival probability of a quasi-energy state $|\langle Z\mu_{i6},\theta_6|\mathbf{U}_n(\theta_6)|Z\mu_{i6},\theta_6\rangle|^2$ for the orbit $e=6$ during 12 almost-periods with $\frac{\omega_0}{\omega_1} = 3.4$. The circles show the survival probabilities at each almost-period $\{mp_{\epsilon,\theta_6}\}_{m=0,...,12}$.}
\end{figure}

\begin{figure}
  \includegraphics[width=9cm]{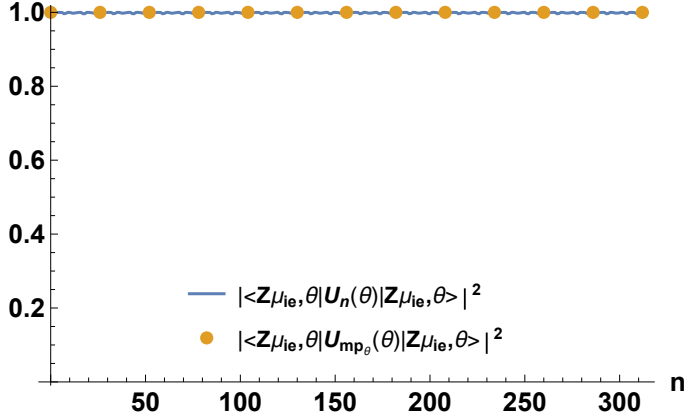}
  \caption{\label{survprobZmu2} Same as fig. \ref{survprobZmu} but with $\frac{\omega_1}{\omega_0} = 0.04$.}
\end{figure}

\begin{figure}
  \includegraphics[width=9cm]{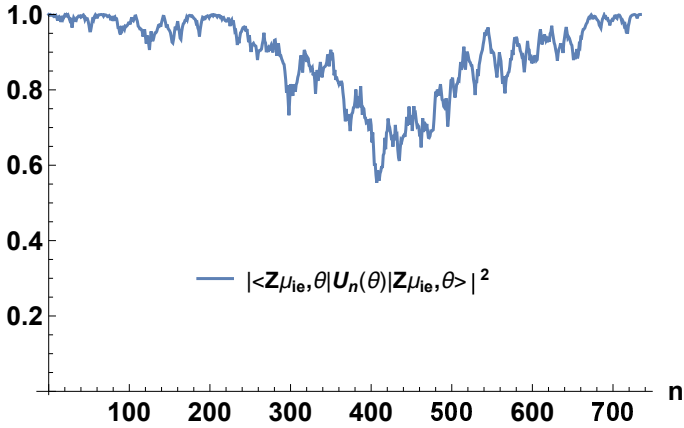}
  \caption{\label{probsurvchaos} Survival probability of a quasi-energy state $|\langle Z\mu_{i0},\theta_0|\mathbf{U}_n(\theta_0)|Z\mu_{i0},\theta_0\rangle|^2$ for the chaotic orbit $e=0$ during 1 almost-period with $\frac{\omega_1}{\omega_0} = 0.03$.}
\end{figure}

By construction, we know that $|\langle Z\mu_{ie},\varphi^{n+1}(\theta)|\mathbf{U}_n(\theta)|Z\mu_i,\theta\rangle|^2=1$. We are interested by the survival probability $P^{surv}_n(\theta) = |\langle Z\mu_{ie},\theta|\mathbf{U}_n(\theta)|Z\mu_{ie},\theta\rangle|^2$, as for example fig. \ref{survprobZmu}.
As expected, the quasi-energy states presents quasi-recurrences at each almost-period (in the same conditions than the previous discussion about the stroboscopic fidelity). But we can see also, that the fluctuations during the transient regime (between two almost-periods) seems not too large. With different parameters, the quasi-energy state is even an almost steady state as we can see it fig. \ref{survprobZmu2}.
To enlight this behaviour, we have compute the average survival probability:
\begin{equation}
  \overline{P^{surv}(\theta)} = \frac{1}{N+1} \sum_{n=0}^N P^{surv}_n(\theta)
\end{equation}
during $N=120$ almost-periods for the three regimes (low, medium and high frequency). The results are presented table \ref{SurvProb}.
\begin{table}
  \caption{\label{SurvProb} Average survival probability of a quasi-energy state during 120 almost-periods for different orbits and different values of $\frac{\omega_1}{\omega_0}$.  For the orbit $e=0$ the presented results correspond to $\epsilon=10^{-1}$ (whereas $\epsilon=10^{-2}$ for the other orbits). We writte in bold the good results ($\geq 97\%$), in sans serif style the correct results ($\geq 90\%$), in normal style the middling results ($\geq 75\%$) and in italic the bad results ($<75\%$).}
  \scriptsize
  \begin{tabular}{|c|c|c|c|c|c|c|c|c|c|}
    \hline
    $e$ & \multicolumn{3}{|c|}{High freq. $\frac{\omega_1}{\omega_0}\ll 1$} & \multicolumn{3}{|c|}{Medium freq. $\frac{\omega_1}{\omega_0} \sim 1$} & \multicolumn{3}{|c|}{Low freq. $\frac{\omega_1}{\omega_0}\gg 1$} \\
    \cline{2-10}
     & $\frac{\sqrt{2}}{100}$ & $0.03$ & $0.04$ & $\sqrt{2}$ & $3.4$ & $4.5$ & $100\sqrt{2}$ & $101.3$ & $104.5$ \\
    \hline
    0 & $\mathit{42.6\%}$ & $\mathit{54.7\%}$ & $\mathit{48.0\%}$ & $\mathit{46.3\%}$ & $\mathit{49.6\%}$ & $\mathit{49.4\%}$ & $\mathit{50.0\%}$ & $\mathit{52.0\%}$ & $\mathit{51.3\%}$ \\
    \hline
    1 & $\mathsf{93.7\%}$ & $\mathsf{94.4\%}$ & $80.0\%$ & $\mathsf{92.2\%}$ & $\mathbf{97.0\%}$ & $\mathbf{99.6\%}$ & $\mathsf{94.9\%}$ & $\mathit{72.1\%}$ & $89.\%$ \\
    \hline
    2 & $\mathbf{99.9\%}$ & $\mathbf{99.9\%}$ & $\mathbf{99.8\%}$ & $\mathsf{90.5\%}$ & $87.7\%$ & $82.5\%$ & $80.7\%$ & $\mathit{45.0\%}$ & $\mathit{55.1\%}$ \\
    \hline
    3 & $\mathbf{99.9\%}$ & $\mathbf{99.8\%}$ & $\mathbf{99.8\%}$ & $88.7\%$ & $\mathsf{91.7\%}$ & $\mathsf{95.9\%}$ & $\mathit{62.8\%}$ & $\mathsf{95.7\%}$ & $\mathit{58.3\%}$ \\
    \hline
    4 & $\mathbf{99.9\%}$ & $\mathbf{99.8\%}$ & $\mathbf{99.8\%}$ & $\mathbf{99.3\%}$ & $\mathbf{97.6\%}$ & $84.5\%$ & $\mathbf{97.3\%}$ & $\mathsf{95.4\%}$ & $\mathbf{98.2\%}$ \\
    \hline
    5 & $\mathbf{99.9\%}$ & $\mathbf{100\%}$ & $\mathbf{100\%}$ & $\mathbf{99.7\%}$ & $\mathbf{98.0\%}$ & $\mathbf{98.4\%}$ & $87.9\%$ & $\mathbf{99.0\%}$ & $75.0\%$ \\
    \hline
    6 & $\mathbf{99.9\%}$ & $\mathbf{99.9\%}$ & $\mathbf{99.9\%}$ & $\mathbf{99.3\%}$ & $88.8\%$ & $\mathbf{98.0\%}$ & $\mathit{68.9\%}$ & $\mathsf{96.8\%}$ & $\mathsf{94.8\%}$ \\
    \hline
    7 & $\mathbf{99.9\%}$ & $\mathbf{99.9\%}$ & $\mathbf{99.9\%}$ & $\mathbf{98.9\%}$ & $\mathit{67.5\%}$ & $\mathbf{97.0\%}$ & $\mathit{55.9\%}$ & $80.3\%$ & $\mathsf{92.6\%}$ \\
    \hline
    8 & $\mathbf{100\%}$ & $\mathbf{100\%}$ & $\mathbf{100\%}$ & $\mathbf{100\%}$ & $\mathbf{100\%}$ & $\mathbf{99.9\%}$ & $\mathbf{98.2\%}$ & $\mathsf{98.0\%}$ & $\mathbf{99.8\%}$ \\
    \hline
  \end{tabular}
\end{table}
The steadiness of the quasi-energy state seems better in the center of the islands, confirming that $\angle \Orb(\theta)$ must be small. Moreover it seems better for orbits with small almost-periods. Due to the dependency to $\frac{\omega_1}{\omega_0}$ of $V(\theta)$, the steadiness of the quasi-energy state is valid only for high and medium frequency regimes. Note that the results do not depend on the time scale, we find very close survival probabilities by averaging during 12 almost-periods and by averaging during 120 almost-periods.\\
For the chaotic orbit $e=0$, we can have an almost steadiness of the quasi-energy states only during the first almost-period (because of the sensitivity to initial conditions), and for $\epsilon$ not too small (because $p_{\epsilon,\theta_0}$ must not be too large). We can see this with fig. \ref{probsurvchaos} and table \ref{probsurvchaos2}.
\begin{table}
  \caption{\label{probsurvchaos2} Same as table \ref{SurvProb} but with the averaging during 1 almost-period and with $\epsilon=10^{-1}$.}
    \scriptsize
   \begin{tabular}{|c|c|c|c|c|c|c|c|c|c|}
    \hline
    $e$ & \multicolumn{3}{|c|}{High freq. $\frac{\omega_1}{\omega_0}\ll 1$} & \multicolumn{3}{|c|}{Medium freq. $\frac{\omega_1}{\omega_0} \sim 1$} & \multicolumn{3}{|c|}{Low freq. $\frac{\omega_1}{\omega_0}\gg 1$} \\
    \cline{2-10}
     & $\frac{\sqrt{2}}{100}$ & $0.03$ & $0.04$ & $\sqrt{2}$ & $3.4$ & $4.5$ & $100\sqrt{2}$ & $101.3$ & $104.5$ \\
    \hline
    0 & $89.7\%$ & ${89.3.7\%}$ & $\mathsf{96.8\%}$ & $\mathit{57.4\%}$ & $\mathit{73.3\%}$ & $\mathit{47.3\%}$ & ${75.7\%}$ & $\mathit{70.3\%}$ & $\mathit{53.9\%}$ \\
    \hline
   \end{tabular}
\end{table}
But since the $\Orb(\theta_0)$ covers a large part of $\mathbb T^2$, the steadiness is middling for the chaotic orbit.\\

As shown in ref. \cite{Guerin1, Guerin2}, in semi-classical light-matter interaction, the Floquet quasienergy states associated with the periodic oscillations of the electromagnetic field are related to the quantum dressed states (of the atom dressed by the photons). So in the semi-classical theory, the Floquet quasienergy states can be viewed as the states of the atom dressed by the classical electromagnetic field. By extension, we can view the SK quasienergy states $|Z\mu_{ie},\theta\rangle$ as the states of the spin (in this example) dressed by the chaotic environment described by the standard map (i.e. the spin dressed by the the ``set'' of ultrashort magnetic pulses). The dressed picture consists to consider a larger system constituted by the quantum system and its ``environment'' described by a classical flow. So an eigenvector of the dressed system physically corresponds to an equilibrium between the system and its environment. For the periodic cases, this equilibrium is a true steady state. But for the almost-periodic cases, due to the irregularities of the almost-periodicity, a quasienergy state represents a dynamical equilibrium rather than a steady equilibrium. As viewed fig. \ref{survprobZmu} with small irregularities (island of stability), the quantum system can temporalily leaves the quasienergy state, but the dynamics send the system back to it, inducing the small fluctuations. The main regularity, i.e. the almost-periodicity, ensures only the quasi-recurrence at each almost-period. And as viewed fig. \ref{survprobZmu2}, at high frequency, the spin being kicked very quickly, it does not have time to change between two kicks. The steadiness is then better. In contrast, in the chaotic sea (fig. \ref{probsurvchaos}), the irregularities are large and the ``return to equilibrium'' of the dynamics starting from a quasienergy state, is difficult. This induces large erratic fluctuations.

\subsection{True steady states}
\begin{figure}
  \includegraphics[width=9cm]{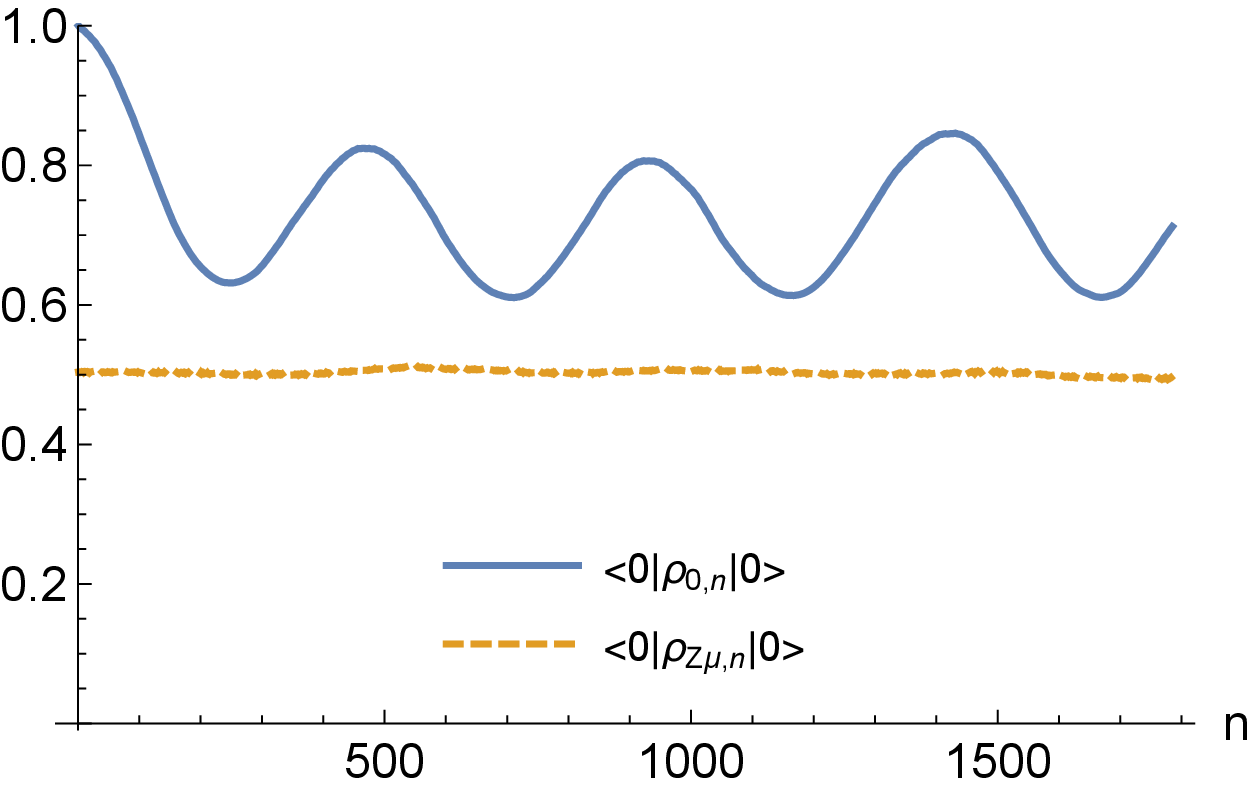}\\
  \includegraphics[width=9cm]{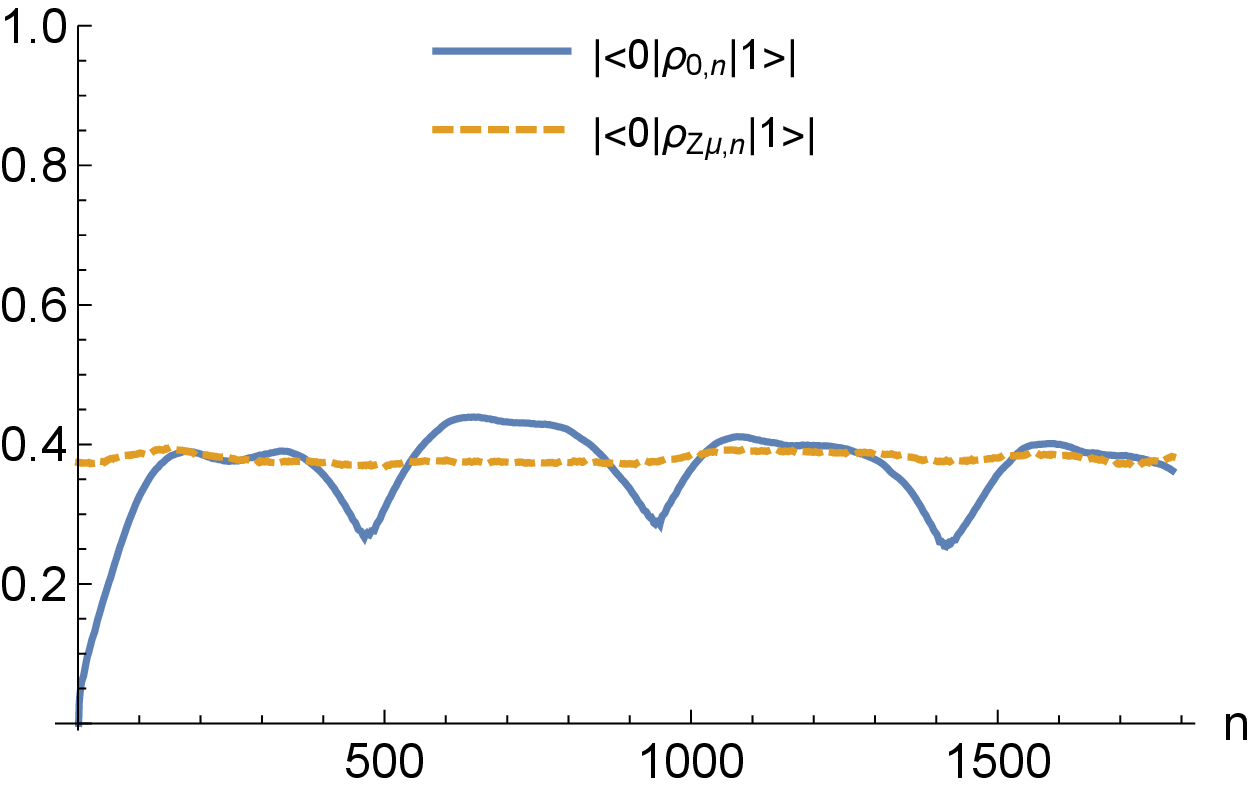}
  \caption{\label{stat_island} Population of the state $|0\rangle$ (up) and coherence (down) of the mixed states $\rho_{0,n}$ (with $\rho_{0,0} = |0\rangle \langle 0|$) and $\rho_{Z\mu,n}$ (with $\rho_{Z\mu,0} = \frac{1}{N} \sum_{m=1}^N |Z\mu_{ie_{(m)}},\theta_{(m)} \rangle \langle Z\mu_{ie_{(m)}},\theta_{(m)} |$) during 2 almost-periods in the chaotic sea (with $\epsilon = 2\times 10^{-1}$) with $\frac{\omega_0}{\omega_1}=3.4$}
\end{figure}

The steadiness viewed in the previous section is just an approximation in some conditions. But we can use eq. \ref{SKquasistate} to built true steady states by considering a set of copies of the driven quantum system and not only a single one. For example, consider $\Orb(\theta_6)$ and 26 copies of the driven quantum system. We suppose that the copy labeled by $(m)$ ($m \leq 25$), has $\varphi^m(\theta_6)$ as initial condition for its kick system. We choose as initial states for the different copies, the quasi-energy states:
\begin{equation}
  |\psi^{(m)}_0 \rangle = |Z\mu_{i6},\varphi^m(\theta_6) \rangle
\end{equation}
After one quick, we have
\begin{equation}
  U(\varphi^m(\theta_6))|\psi^{(m)}_0 \rangle = e^{-\imath \chi_{i6}}|Z\mu_{i6},\varphi^{m+1}(\theta_6) \rangle
\end{equation}
Everything happens as if the copy $(m)$ takes the place of the copy $(m+1)$ (for $m<25$) and as if the copy $(25)$ takes the place of the copy $(0)$ (since $\varphi^{26}(\theta_6) = \theta_6 + \mathcal O(\epsilon)$). The set of 26 copies is then unchanged by the dynamics (even if individually each copy changes). We have then a true steady state (up to an error of magnitude $\epsilon$) by considering the mixed state of the set of copies:
\begin{equation}
  \rho_n = \frac{1}{26} \sum_{m=0}^{25} \mathbf{U}_n(\varphi^{m}(\theta_6)) |\psi^{(m)}_0 \rangle \langle \psi^{(m)}_0 | \mathbf{U}_n(\varphi^{m}(\theta_6))^\dagger
\end{equation}
If the initial mixed state is the quasi-energy state of the orbit, $\rho_n$ does not evolve up to an error of magnitude $\epsilon$. This reasoning can be applied to all orbit, and to many orbits together. For example, we consider the small island of stability into the chaotic sea. We consider $N=1391$ copies of the quantum system, with kick initial conditions corresponding to the points of $\Orb(\theta_6)$, $\Orb(\theta_7)$, $\Orb(\theta_8)$ and the 893 points of $\Orb(\theta_0)$ corresponding to the precision $\epsilon = 2\times 10^{-1}$ ($\epsilon=10^{-2}$ for the orbits in the small island). We consider the following density matrix
\begin{equation}
  \rho_n = \frac{1}{N} \sum_{m=1}^N \mathbf{U}_n(\theta_{(m)}) |\psi^{(m)}_0 \rangle \langle \psi^{(m)}_0| \mathbf{U}_n(\theta_{(m)})^\dagger
\end{equation}
where $\theta_{(m)}$ is the kick initial condition of the copy $(m)$; with the two initial states: $|\psi^{(m)}_0 \rangle = |Z\mu_{ie_m},\theta_{(m)}\rangle$ (with $\theta_{(m)} \in \Orb(\theta_{e_m})$), and $|\psi^{(m)}_0 \rangle = |0\rangle$ (for comparison). A population and the coherence of $\rho_n$ are drawn fig. \ref{stat_island}.
As expected, $\rho_{Z\mu,0}$ is a steady states whereas any state presents oscillations due to the almost-periodicities of the four orbits.\\

We can relate this result to another physical situation. Suppose that we sent a regular train of ultrashort magnetic pulses onto a spin lattice (constituted by the $N=1391$ spins in a solid medium without mutual interaction). The pulses must then go accross the matter which is the support of the spins before reaching these ones, the scattering of light in the medium being supposed chaotic. We model this chaotic scattering by the standard map: it induces delays and polarization changes on the scattered pulses. Each spin views then a different train described by a different initial condition $\theta_{(m)}$. The spins associated with $\theta_{(m)}$ in the double small island (almost regular orbits $\{\Orb(\theta_6),\Orb(\theta_7),\Orb(\theta_8)\}$) are close to the surface of the solid (the pulses go accross a small distance and the scattering is still regular), whereas the spins associated with $\theta_{(m)}$ in the chaotic sea are deep in the solid (the scattering of the pulses is now chaotic). Due to the chaotic scattering, the spin lattice is submitted to decoherence and relaxation processes. Its density matrix evolves then to a steady state which is actually $\rho_{Z\mu,0}$ (theorem 5 in ref. \cite{Viennot3} proves that the mixed state of such a system converges at long time to a combination of fundamental quasienergy density matrices).

\section{Conclusion}
The effective Hamiltonian defined by the Schr\"odinger-Koopman approach permits to extend the approach of the Floquet effective Hamiltonian to quasi-periodically driven systems without knowledge of the different frequencies of these systems. It can be also applied to chaotically driven systems, whereas in  this case it is diffult to exploit the dynamical behaviour (but this a direct consequence of the definition of chaos). 

\ack
The author acknowledge support from ISITE Bourgogne-Franche-Comt\'e (contract ANR-15-IDEX-0003) under grants from I-QUINS and GNETWORKS projects, and support from the R\'egion Bourgogne-Franche-Comt\'e under grants from APEX project. Numerical computations have been executed on computers of the Utinam Institute supported by the R\'egion Bourgogne-Franche-Comt\'e and the Institut des Sciences de l'Univers (INSU).\\

The author thanks Professor Atushi Tanaka for useful exchanges.

\appendix
\section{BCH formula} \label{BCH}
For some computations, we consider the following version of the Baker-Campbell-Hausdorff formula \cite{Schmid}:
\begin{equation}
  e^X e^Y = e^{X + f_X[Y] + \mathcal O(\|Y\|^2)}
\end{equation}
where $f_X = f(\ad_X)$ with $\ad_X[Y] = [X,Y]$ ($\ad_X$ is an operator onto the bounded operator space $\mathcal B(\mathcal H)$) and $f(x) = \frac{x}{1-e^{-x}}$ ($f(\ad_X)$ is defined by functional calculus \cite{RS}). We have also
\begin{equation}
  e^Y e^X = (e^{-X}e^{-Y})^{-1} = e^{X + f_{-X}[Y] +  \mathcal O(\|Y\|^2)}
\end{equation}
and conversely
\begin{equation}
  e^{X+Y} = e^{f^{-1}_{-X}[Y] + \mathcal O(\|Y\|^2)} e^X
\end{equation}

To compute $f_X$ we have two possibilities. The first one consists to use the Taylor serie of $f$:
\begin{equation}
  f(\ad_X) = 1 - \sum_{n=1}^{+\infty} \frac{(1-\Ad(e^X))^n}{n(n+1)}
\end{equation}
where $\Ad(g)Y=gYg^{-1}$. The second one consists to consider the Hilbert-Schmidt space of the operators of $\mathcal H$ (\cite{RS}). To simplify the discussion here, we suppose that $\mathcal H$ is finite dimensional ($\dim \mathcal H=N$). The Hilbert-Schmidt space can be identified to $\mathcal{HS} = \mathbb C^{N^2}$. The Hilbert-Schmidt representation of an operator $Y \in \mathcal L(\mathcal H)$ is then
\begin{equation}
  Y = \left(\begin{array}{ccc} Y_{11} & ... & Y_{1N} \\ \vdots & \ddots & \vdots \\ Y_{N1} & ... & Y_{NN} \end{array} \right) \to |Y \rrangle = \left(\begin{array}{c} Y_{11} \\ \vdots \\ Y_{1N} \\ Y_{21} \\ \vdots \\ Y_{NN} \end{array} \right)
\end{equation}
for a matrix representation in the choosen orthonormal basis. The inner product of $\mathcal{HS}$ is $\llangle Z|Y\rrangle = \mathrm{tr}(Z^\dagger Y)$. We have moreover
\begin{eqnarray}
  |XY\rrangle & = & X \otimes 1_N |Y\rrangle \\
  |YX\rrangle & = & 1_N \otimes X^T |Y \rrangle
\end{eqnarray}
It follows that $\ad_X = X \otimes 1_N - 1_N \otimes X^T$ (where $\phantom{X}^T$ denotes the transposition). $\ad_X$ can be then viewed as a $N^2$-order square matrix. Let $\{x_n\}_{n=1,...,N^2}$ be the spectrum of $\ad_X$ and $P$ be such that $P^{-1}\ad_X P = \mathrm{diag}(x_1,...,x_{N^2})$ (the diagonal matrix having $(x_1,...,x_{N^2})$ on the diagonal). We have then $f(\ad_X) = P \mathrm{diag}(f(x_1),...,f(x_{N^2})) P^{-1}$.\\

For example to find eq. (\ref{relatHeff}) we start from $e^{A_{e\nu}\tilde \epsilon^\nu + \mathcal O(\epsilon^2) } e^{-\imath p H^{eff}} = e^{-\imath p H^{eff} + f_{\imath p H^{eff}}[A_{e\nu}] \tilde \epsilon^\nu + \mathcal O(\epsilon^2)}$. $H^{eff} = \mathrm{diag}(\chi_{e1},...,\chi_{eN})$ in the basis $(|Z\mu_i,\theta\rangle)_{i=1,...,N}$, and then $\ad_{H^{eff}} = \mathrm{diag}(0,\chi_{e1}-\chi_{e2},...,\chi_{e1}-\chi_{eN},\chi_{e2}-\chi_{e1},0,...,\chi_{e2}-\chi_{eN},...,\chi_{eN}-\chi_{eN-1},0)$. It follows that
\begin{equation}
  f(\ad_{\imath p H^{eff}}) = \mathrm{diag} \left( \left\{\begin{array}{ll} 1 & \text{if $j=i$} \\ \frac{\imath p(\chi_{ei}-\chi_{ej})}{1-e^{-\imath p(\chi_{ei}-\chi_{ej})}} & \text{if $j\not=i$} \end{array}\right. \right)_{i,j}
\end{equation}
($\lim_{x \to 0} f(x)=1$).

\section{Direct computation of the effective Hamiltonian}\label{HKcomp}
Eq. \ref{SKquasistate} which defines the quasienergy states can be rewritten as follows:
\begin{equation}
  U_K|Z\mu_{ie} \rrangle = e^{-\imath \chi_{ie}} |Z\mu_{ie} \rrangle
\end{equation}
where $|Z\mu_{ie} \rrangle \in \mathcal H \otimes L^2(\Gamma,d\mu)$ and $U_K = \mathcal T^{-1} U$, with $\langle \theta|Z\mu_{ie} \rrangle = |Z\mu_{ie},\theta \rangle \in \mathcal H$, $\langle \theta'|U|\theta \rangle = U(\theta) \delta(\theta-\theta')$, and $\forall \zeta \in L^2(\Gamma,d\mu)$
\begin{equation}
  (\mathcal T^{-1} \zeta)(\theta) = \zeta(\varphi^{-1}(\theta))
\end{equation}
or in other words $\mathcal T^{-1} = \int_\Gamma |\varphi^{-1}(\theta)\rangle \langle \theta| d\mu(\theta)$. We have then $U_K = e^{-\imath H_K}$ with
\begin{equation}
  H_K = \sum_e \sum_i \sum_\lambda (\chi_{ie}-\imath \lambda) f_\lambda |Z\mu_{ie} \rrangle \llangle Z\mu_{ie} | f_\lambda^*
\end{equation}
by taking into account of the gauge changes ($\mathcal T f_\lambda = e^\lambda f_\lambda$). We can then write that:
\begin{equation}
  H^{eff}(\theta) = \langle \theta|P_0 H_K P_0 | \theta \rangle
\end{equation}
where $P_0 \cdot P_0$ is a spectral filtering selecting only fundamental quasienergies (i.e. excluding the redundant quasienergies associated with the gauge changes).\\
Concretely, let $\Gamma_e$ be an ergodic component of $\Gamma$. Let $\Orb_\epsilon(\theta_0)= \{\theta_n = \varphi^n(\theta_0) \}_{n \in \{0,...,p_{\epsilon,\theta_0}-1\}}$ be the approximated orbit of $\theta_0$. By definition, $\|\varphi^{p_{\epsilon,\theta_0}}(\theta_0) - \theta_0 \|= \mathcal O(\epsilon)$ and since $\overline{\Orb(\theta_0)}=\Gamma_e$, we have $\forall \theta \in \Gamma_e$, $\exists n \in \{0,...,p_{\epsilon,\theta_0}-1\}$ such that $\|\theta-\theta_n\| = \mathcal O(\epsilon)$. $\Orb_\epsilon(\theta_0)$ can be then viewed as a partition (a discrete approximation) of $\Gamma_e$. The Hilbert space generated by $(|\theta_n\rangle)_{n \in \{0,...,p_{\epsilon,\theta_0}-1\}}$ is then a finite representation of $L^2(\Gamma_e,d\mu)$ with the accuracy $\epsilon$. The restriction to $\Gamma_e$ of the Koopman operator $\mathcal T^{-1}$ can then represented by
\begin{eqnarray}
  \mathcal T^{-1}_{|\Gamma_e} \simeq \mathfrak T_e^{-1} & = & \sum_{n=1}^{p_{\epsilon,\theta_0}-1} |\theta_n\rangle \langle \theta_{n-1}| + |\theta_0\rangle \langle \theta_{p_{\epsilon,\theta_0}-1}| \\
   & = & \left(\begin{array}{ccccc} 0 & 0  & ... & 0 &  1 \\ 1 & 0  & ... & 0 & 0 \\  0 & 1  &  & 0 & 0 \\\vdots & \vdots & \ddots & & \vdots \\ 0 & 0  & ... & 1 & 0 \end{array} \right)
\end{eqnarray}
It follows that $U_K$ restricted to $\Gamma_e$ can be approximated by the $(\dim \mathcal H \times p_{\epsilon,\theta_0})$-order matrix
\begin{eqnarray}
  \mathfrak U_{Ke} & = & \sum_{n=0}^{p_{\epsilon,\theta_0}-2} U(\theta_n)\otimes |\theta_{n+1}\rangle \langle \theta_n| \nonumber \\
  & & \quad + U(\theta_{p_{\epsilon,\theta_0}-1}) \otimes |\theta_0\rangle \langle \theta_{p_{\epsilon,\theta_0}-1}|
\end{eqnarray}
Let $(e^{-\imath \chi_{ae}})_{a \in \{1,...,p_{\epsilon,\theta_0}\dim \mathcal H\}} = \Sp(\mathfrak U_{Ke})$ be the spectrum of $\mathfrak U_{Ke}$ and $(|Z\mu_{ae} \rrangle \in \mathbb C^{p_{\epsilon,\theta_0} \dim \mathcal H})_{n \in \{1,...,p_{\epsilon,\theta_0}\dim \mathcal H\}}$ be the associated eigenvectors. The spectrum of $\mathfrak T_e^{-1}$ is constituted by the $p_{\epsilon,\theta_0}$-th roots of unity: $\Sp(\mathfrak T_e^{-1}) = \left(e^{\imath \frac{2n\pi}{p_{\epsilon,\theta_0}}}\right)_{n \in \{0,...,p_{\epsilon,\theta_0}-1\}}$. The spectral filtering consists then to select $\dim \mathcal H$ eigenvalues $(e^{\imath \chi_{ie}})_{i \in I_e}$ (with $\mathrm{card} I_e = \dim \mathcal H$) such that:
\begin{equation}
\forall i,j\in I_e, i\not=j, \quad (\chi_{ie} - \chi_{je} \mod 2\pi) \not\in \frac{2 \mathbb Z \pi}{p_{\epsilon,\theta_0}}
\end{equation}
finally, we have $\forall \theta \in \Gamma_e$
\begin{equation}
  H^{eff}(\theta) = \sum_{i\in I_e} \chi_{ie} \langle \theta_n|Z\mu_{ie} \rrangle \llangle Z\mu_{ie}|\theta_n\rangle + \mathcal O(\epsilon)
\end{equation}
with $\theta_n$ such that $\|\theta-\theta_n\| = \mathcal O(\epsilon)$, $\langle \theta_n|Z\mu_{ie} \rrangle \in \mathbb C^{\dim \mathcal H}$ being the (renormalized) vector extracted from $|Z\mu_{ie} \rrangle \in \mathbb C^{p_{\epsilon,\theta_0} \dim \mathcal H}$ by choosing the $\dim \mathcal H$ components associated with the $(n+1)$-th vector of the basis of $\mathbb C^{p_{\epsilon,\theta_0}}$.\\
The main difficulty with this method is that it needs the diagonalization of the large matrix $\mathfrak U_{Ke}$ whereas we need only $\dim \mathcal H$ eigenvectors. For very large matrices, we can compute only the needed $\dim \mathcal H$ eigenvectors by a matrix partitioning method \cite{Killingbeck}.


\section*{References}
\begin{thebibliography}{0}
\bibitem{Floquet} Floquet G 1883 {\it Ann. \'Ecole Normale Sup\'erieure} {\bf 12}, 47.
\bibitem{Shirley} Shirley J H 1965 {\it Phys. Rev.} {\bf 138}, B979.
\bibitem{Sambe} Sambe H 1973 {\it Phys. Rev. A} {\bf 7} 2203.
\bibitem{Barone} Barone S R and Narcowich M A 1977 {\it Phys. Rev. A} {\bf 15}, 1109.
\bibitem{Moore1} Moore D J and Stefman G E 1990 {\it J. Phys. A: Math. Gen.} {\bf 23}, 2049.
\bibitem{Haake} Haake F 1991 {\it Quantum Signatures of Chaos} (Springer: Berlin).
\bibitem{Moore2} Moore D J and Stedman G E 1992 {\it Phys. Rev. A} {\bf 45}, 513.
\bibitem{Guerin1} Gu\'erin S 1997 {\it Phys. Rev. A} {\bf 56}, 1458.
\bibitem{Guerin2} Gu\'erin S, Monti F, Dupont J M and Jauslin H R 1997 {\it J. Phys. A: Math. Gen.} {\bf 30}, 7193.
\bibitem{Drese} Drese K and Holthaus M 1999 {\it Eur. Phys. J. D} {\bf 5}, 119.
\bibitem{Guerin3} Gu\'erin S and Jauslin H R 2003 {\it Adv. Chem. Phys.} {\bf 125}, 147.
\bibitem{Miyamoto} Miyamoto M and Tanaka A 2007 {\it Phys. Rev. A} {\bf 76}, 042115.
\bibitem{Viennot1} Viennot D 2009 {\it J. Phys. A: Math. Theor.} {\bf 42}, 395302.
\bibitem{Else} Else D V, Bauer B and Nayak C 2016 {\it Phys. Rev. Lett.} {\bf 117}, 090402.
\bibitem{Goldman} Goldman N and Dalibard J 2014 {\it Phys. Rev. X} {\bf 4}, 031027.
\bibitem{Neumann} Neumann E and Pikovsky A 2002 {\it Eur. Phys. J. D} {\bf 26}, 219.
\bibitem{Verdeny1} Verdeny A, Rudnicki L, M\"uller C A and Mintert F 2014 {\it Phys. Rev. Lett.} {\bf 113}, 010501.
\bibitem{Verdeny2} Verdeny A, Puig J and Mintert F 2016 {\it Z. Naturforsch.} {\bf 71}, 897.
\bibitem{Lasota} Lasota A and Mackey M C 1994, {\it Chaos, Fractals and Noise} (Springer: New York).
\bibitem{Viennot2} Viennot D and Aubourg L 2013 {\it Phys. Rev. E} {\bf 87}, 062903.
\bibitem{Koopman1} Koopman B O 1931 {\it Proc. Natl. Acad. Sci. USA} {\bf 17}, 315.
\bibitem{Koopman2} Koopman B O and von Neumann J 1932 {\it Proc. Natl. Acad. Sci. USA} {\bf 18}, 255.
\bibitem{Viennot3} Viennot D and Aubourg L 2018 {\it J. Phys. A: Math. Theor.} {\bf 51}, 335201.
\bibitem{Bocchieri} Bocchieri P and Loinger A 1957 {\it Phys. Rev.} {\bf 107}, 337.
\bibitem{Eberly} Eberly J H, Narozhny N B and Sanchez-Mondragon J J 1980 {\it Phys. Rev. Lett.} {\bf 44}, 1323.
\bibitem{Chirikov} Chirikov B V 1979 {\it Phys. Rep.} {\bf 52}, 263.
\bibitem{Poletti} Poletti D and Kollath C 2011 {\it Phys. Rev. A} {\bf 84}, 013615.
\bibitem{Schmid} Schmid W 1982 {\it Bull. Amer. Math. Soc.} {\bf 6}, 175.
\bibitem{RS} Reed M and Simon B 1980 {\it Methods of modern mathematical physics I - functional analysis} (Academic Press: New York).
\bibitem{Killingbeck} Killingbeck J P and Jolicard G 2003 {\it J. Phys. A: Math. Gen.} {\bf 36}, R105 (2003). 
\end{thebibliography}
\end{document}